\documentclass[twocolumn]{aastex62}
\usepackage{hyperref}
\usepackage{amsmath}
\usepackage{amssymb}
\usepackage{graphicx}
\usepackage[utf8]{inputenc}

\DeclareRobustCommand{\ion}[2]{\textup{#1\,\textsc{\lowercase{#2}}}}

\newcommand\fei{\ion{Fe}{i}}
\newcommand\feii{\ion{Fe}{ii}}
\newcommand\srii{\ion{Sr}{ii}}
\newcommand\euii{\ion{Eu}{ii}}
\newcommand\baii{\ion{Ba}{ii}}
\newcommand\tii{\ion{Ti}{i}}
\newcommand\tiii{\ion{Ti}{ii}}
\newcommand\oi{\ion{O}{i}}
\newcommand\nai{\ion{Na}{i}}
\newcommand\mgi{\ion{Mg}{i}}
\newcommand\ali{\ion{Al}{i}}
\newcommand\cai{\ion{Ca}{i}}
\newcommand\scii{\ion{Sc}{ii}}
\newcommand\vi{\ion{V}{i}}
\newcommand\vii{\ion{V}{ii}}
\newcommand\cri{\ion{Cr}{i}}
\newcommand\crii{\ion{Cr}{ii}}
\newcommand\mni{\ion{Mn}{i}}
\newcommand\coi{\ion{Co}{i}}
\newcommand\nii{\ion{Ni}{i}}
\newcommand\zni{\ion{Zn}{i}}
\newcommand\ki{\ion{K}{i}}
\newcommand\sii{\ion{Si}{i}}

\newcommand{\kms}{km\,s$^{-1}$}
\newcommand{\teff}{$T_{\rm eff}$}
\newcommand{\logg}{$\log g$}
\newcommand{\vt}{$\xi_{t}$}

\newcommand{\rproc}{$r$-process}
\newcommand{\sproc}{$s$-process}

\newcommand{\AB}[2]{$\mbox{[#1/#2]}$}
\newcommand{\feh}{\AB{Fe}{H}}

\newcommand{\rs}{$r/s$}
\newcommand{\rI}{$r$-I}
\newcommand{\rII}{$r$-II}
\newcommand{\limr}{limited-$r$}

\usepackage[utf8]{inputenc}

\shorttitle{The $R$-process Alliance : First Magellan/MIKE $r$-process-enhanced star search release}
\shortauthors{Ezzeddine et al.}
\begin{document}

\title{The $R$-process Alliance : First Magellan/MIKE Release from the Southern Search for $R$-Process-enhanced Stars\footnote{This paper includes data gathered with the 6.5 meter Magellan Telescopes located at Las Campanas Observatory, Chile.}}                         

\newcommand{\alex}[1]{\textcolor{orange}{(APJ: #1)}}
\newcommand{\alexs}[2]{\textcolor{orange}{(APJ: \sout{#1} #2)}}

\correspondingauthor{Rana Ezzeddine }
\email{ranae@mit.edu}

\author{Rana Ezzeddine}

\affiliation{Department of Astronomy, University of Florida, Bryant Space Science Center, Gainesville, FL 32611, USA}
\affiliation{Joint Institute for Nuclear Astrophysics - Center for Evolution of the Elements, USA}
\affiliation{Department of Physics \& Kavli Institute for Astrophysics and Space Research, Massachusetts Institute of Technology, Cambridge, MA 02139, USA}

\author{Kaitlin Rasmussen}

\affiliation{Department of Physics, University of Notre Dame, Notre Dame, IN 46556, USA}
\affiliation{Joint Institute for Nuclear Astrophysics - Center for Evolution of the Elements, USA}  

\author{Anna Frebel}
\affiliation{Department of Physics \& Kavli Institute for Astrophysics and Space Research, Massachusetts Institute of Technology, Cambridge, MA 02139, USA}
\affiliation{Joint Institute for Nuclear Astrophysics - Center for Evolution of the Elements, USA}

\author{Anirudh Chiti}
\affiliation{Department of Physics \& Kavli Institute for Astrophysics and Space Research, Massachusetts Institute of Technology, Cambridge, MA 02139, USA}
\affiliation{Joint Institute for Nuclear Astrophysics - Center for Evolution of the Elements, USA}

\author{Karina Hinojisa}
\affiliation{Department of Physics \& Kavli Institute for Astrophysics and Space Research, Massachusetts Institute of Technology, Cambridge, MA 02139, USA}

\author{Vinicius M. Placco}
\affiliation{Department of Physics, University of Notre Dame, Notre Dame, IN 46556, USA}
\affiliation{Joint Institute for Nuclear Astrophysics - Center for Evolution of the Elements, USA}

\author{Ian U. Roederer}
\affiliation{Department of Astronomy, University of Michigan, Ann Arbor, MI 48109, USA}
\affiliation{Joint Institute for Nuclear Astrophysics - Center for Evolution of the Elements, USA}

\author{Alexander P. Ji}
\affiliation{The Observatories of the Carnegie Institution for Science, 813 Santa Barbara St., Pasadena, CA 91101, USA}
\affiliation{Hubble Fellow}
\affiliation{Joint Institute for Nuclear Astrophysics - Center for Evolution of the Elements, USA}

\author{Timothy C. Beers}
\affiliation{Department of Physics, University of Notre Dame, Notre Dame, IN 46556, USA}
\affiliation{Joint Institute for Nuclear Astrophysics - Center for Evolution of the Elements, USA}

\author{Terese T. Hansen}
\affiliation{Mitchell Institute for Fundamental Physics and Astronomy and Department of Physics and Astronomy, Texas A\&M University, College Station, TX 77843, USA}

\author{Charli M. Sakari}
\affiliation{Department of Astronomy, University of Washington, Seattle, WA 98195-1580, USA}

\author{Jorge Melendez}
\affiliation{Instituto de Astronomia, Geof\'{i}sica e Ci\^{e}ncias Atmosf\'{e}ricas, Universidade de S\~{a}o Paulo, SP 05508-900, Brazil}

\begin{abstract}
Extensive progress has been recently made into our understanding of heavy element production via the \rproc\ in the Universe, specifically with the first observed neutron star binary merger (NSBM) event associated with the gravitational
wave signal detected by LIGO, GW170817. The chemical abundance patterns of metal-poor \rproc-enhanced stars provides key evidence into the dominant site(s) of the \rproc, and whether NSBMs are sufficiently frequent or prolific \rproc\ sources to be responsible for the majority of r-process material in the Universe. We present atmospheric stellar parameters (using a Non-Local Thermodynamic Equilibrium analysis) and abundances from a detailed analysis of 141 metal-poor stars, carried out as part of the $R$-Process Alliance (RPA) effort. We obtained high-resolution ``snapshot" spectroscopy of the stars using the MIKE spectrograph on the 6.5\,m Magellan Clay telescope at Las Campanas Observatory in Chile. 
We find 10 new highly enhanced \rII\ (with [Eu/Fe] $> +1.0$), 62 new moderately enhanced \rI\ ($+0.3 < $ [Eu/Fe] $\le +1.0$) and 17 new \limr\ ([Eu/Fe] $< +0.3$) stars. Among those, we find 17 new carbon-enhanced metal-poor (CEMP) stars, of which five are CEMP-no. We also identify one new $s$-process-enhanced ([Ba/Eu ]$ > +0.5$), and five new \rs\ ($0.0 < $ [Ba/Eu] $ < +0.5$) stars. In the process, we discover a new ultra metal-poor (UMP) star at $\feh=-4.02$.
One of the \rII\ stars shows a deficit in $\alpha$ and Fe-peak elements, typical of dwarf galaxy stars.
Our search for \rproc-enhanced stars by RPA efforts, has already roughly doubled the known \rproc\ sample.  

\end{abstract}

\keywords{nucleosynthesis ---  stars: abundances ---  stars: Population II --- stars: atmospheres --- stars: fundamental parameters}

\section{Introduction}\label{intro}
The production of elements with $Z\ge38$ occurs via the rapid ($r$-) and slow ($s$-) neutron-capture processes. While the site of the $s$-process is 
well-understood \citep{gallino2005,karakas2010,frebel-rev2018}, a number of sites for the \rproc\ have been proposed over the last few decades. A key requirement is the ability to provide the strong neutron flux needed for rapid neutron capture to occur. This has been predicted to happen in extreme sites like neutron star binary mergers (NSBM) (e.g., \citealt{lattimer1974}), magneto-rotationally driven jets (e.g., \citealt{winteler2012}), or collapsar disk winds (e.g., \citealt{siegel2019}).

Metal-poor stars enhanced in \rproc\ elements have been found in both the Milky Way halo (e.g., \citealt{rpa1}, \citealt{rpa2} and \citealt{roederer2018a}) as well as dwarf galaxies, most notably the $r$-process ultra-faint dwarf galaxy Reticulum\,II (Ret\,II) \citep{ji2016Nat,roederer2016}. These stars formed from gas that was previously enriched by a nucleosynthesis event during which heavy $r$-process elements were made. Altogether, they thus provide the best accessible evidence for the early operation of the $r$-process, as many $r$-process elements can be detected in their spectra, and the corresponding chemical abundances be measured. 

In 2017, the LIGO/Virgo gravitational wave observatory discovered the first NSBM event GW170817 \citep{abbott2017}, after which a kilonova, likely powered by the decay of newly synthesized of \rproc\ elements, was observed (e.g., \citealt{pian2017}). Together with chemical-evolution models of the Galaxy that implement NS-NS mergers \citep{cescutti2015,wehmeyer2015,cote2017} and the existence of Reticulum\,II, these lines of evidence support the hypothesis that NS mergers are likely a dominant site of \rproc\ nucleosynthesis. However, improved chemical-evolution models by \cite{cote2018b}, more recently suggested that NSBM may not in fact be the only sources of \rproc\ elements in the Galaxy. Additional sources, such as magneto-rotational (``jet'') supernovae \citep{nishimura2017} could be needed to reproduce the observed \rproc\ patterns in, for e.g., moderately enhanced \rproc\ (\rI: $+0.3<\mathrm{[Eu/Fe]}\leq +1.0$) metal-poor stars. Alternative sites for highly $r$-process-enhanced (\rII: [Eu/Fe] $> +1.0$) stars have also been suggested, such as collapsars \citep{siegel2019}.
This conundrum underscores the need for studying large numbers of \rproc-enhanced stars, to better interpret and place constraints on heavy-element formation through the \rproc, and all of its associated astrophysical site(s) throughout cosmic history. 

Motivated by the importance of \rproc-enhanced halo stars, the $R$-Process Alliance (RPA) collaboration aims at significantly increasing the numbers of such stars, as, until recently, only a relatively small number of them were known. Moreover, a major goal is to combine \rproc\ observations, \rproc\ theory efforts, and results from chemical-evolution simulations to eventually produce a more complete understanding of the origin of the \rproc-enhanced star population in the Galactic halo.  To this end, two extensive ``data release'' papers have already been published by \citet{rpa1} (hereafter RPA-1) and \citet{rpa2} (hereafter RPA-2) which report on many newly discovered \rproc-enhanced stars. This paper (hereafter RPA-3) describes the third data release, based on spectra collected with the 6.5\,m Magellan Clay telescope.

With the emergence of a large \rproc-enhanced star sample, and thus statistically meaningful measures of chemical abundances, kinematics, frequencies, etc. of these stars, questions about their formation and any prior enrichment of their birth gas clouds can soon be addressed more rigorously. This is also of interest with respect to 
establishing possible connections between the \rproc\ signature observed in the Ret\,II stars and that of \rproc\ stars found in the Galactic halo. Given their nearly identical abundance patterns, the halo stars likely originated in ancient dwarf galaxy analogs similar to Ret\,II \citep{roederer2018}. Establishing the origin story of halo $r$-process-enhanced stars will provide important insides into the environment in which the earliest $r$-process events took place.

 From a cosmological perspective, the \rproc-enhanced stars also offer another advantage, since they can be dated from their abundance ratios of long-lived isotopes of Th and U relative to abundances of stable \rproc\ elements such as Eu (e.g., \citealt{cayrel2001,frebel2007,placco2017,holmbeck2018}). With half-lives of 4.5 Gyrs and 14.05 Gyrs, respectively, $^{238}$U and $^{232}$Th decay over cosmologically relevant timescales. 
Such ages could be used, for example, to place lower limits on the time of their associated \rproc\ production event(s). We note, however, that such ratios can not be reliably used for actinide-boost \rproc\ stars \citep{schatz2002}, which are known to exhibit unusually high Th and U abundances as compared to stable elements, such as Eu. Interestingly, the actinide elements ratio, Th/U (if measured simultaneously in a star), seems to remain a robust tool to determine the ages of actinide-boost \rproc\ stars \citep{holmbeck2019}.

This paper is outlined as follows.  In Section\,\ref{sec:obs}, we describe the observations, data reduction, and radial-velocity measurements. In Section\,\ref{sec:stell_param}, we present the stellar parameter determinations of the sample using both 
1D (1-Dimensional), LTE (Local Thermodynamic Equilibrium) and 1D, NLTE (Non-Local Thermodynamic Equilibrium) methods. Section\,\ref{sec:abund} presents the chemical abundances of the light elements and the $\alpha$-elements, as well as the Fe-peak and neutron-capture elements determined for the stars. We then  discuss the results and conclude in Sections\,\ref{sec:disc} and \ref{sec:conc}, respectively.

\section{Observations}\label{sec:obs}
\subsection{Target Selection and Observations}
Metal-poor, candidate \rproc-enhanced target stars were selected from various metal-poor surveys, including the RAdial Velocity Experiment (RAVE; \citealp{RAVE2006}),  Large Sky Area Multi-Object Fibre Spectroscopic Telescope (LAMOST) (C. Liu, private communication), Best \& Brightest (B\&B; \citealp{schlaufman2014,placco2019}), SkyMapper\footnote{The SkyMapper stars are taken from the extremely metal poor (EMP; \feh $<-3$) sample discussed in \citet{dacosta2019}; specifically it includes all stars that satisfy $g_{\mathrm{SkyMapper}}$ $\leq$ 14.1 and --3.0 $\leq$ [Fe/H]$_{\mathrm{fitter}}$ $\leq$ --2.0, where [Fe/H]$_{\mathrm{fitter}}$ is the abundance estimated from the low-resolution spectra (see \citealt{dacosta2019} for details).} \citep{skymapper2018}, Melendez \& Placco (M\&P; \citealp{melendez2016}), and the Hamburg/ESO Survey (HES; \citealp{H-ESO2008}). Most of our targets were selected from the RAVE fourth and fifth data releases (DR4 and DR5; \citealp{kordopatis2013,kunder2017}). For those latter stars, their metal-poor nature was vetted by \citet{placco2018}, who derived estimates of their stellar parameters based on medium-resolution spectroscopy acquired on a number of different telescopes. Additional details on the selection criteria for these stars is presented in \citet{rpa1} and \citet{placco2018,placco2019}.

All targets were observed at high spectral resolution using the Magellan Inamori Kryogenic Echelle (MIKE) spectrograph \citep{bernstein2003} on the Clay (Magellan\,II) telescope at Las Campanas Observatory, Chile. Overall, a total of 148 stars with visual magnitudes ($V$) between 10.0 and 14.5 were observed. The spectra were collected over the course of three years, from January 2015 through December 2018. Either the 0\farcs7 or 1\farcs0 slit with $2\times2$ binning was used, yielding nominal resolving powers of $R\sim$28,000 or $R\sim$22,000 in the red, and $R\sim35,000$ or $R\sim28,000$ in the blue, respectively. All spectra cover the wavelength range from $\sim3860$\,{\AA} to $\sim9000$\,{\AA}. 

Exposure times depended on magnitude and weather conditions, but were typically between 900\,s for brighter stars to 1800\,s for fainter ones, aiming for Signal-to-Noise (S/N) values of 25\,pixel$^{-1}$ at 4000\,{\AA}. When needed, more than one exposure was taken to increase the S/N to the desired level. The quality of these spectra enables the detection or a meaningful upper limit on the \euii\ line at 4129\,{\AA}, which serves as the major \rproc\ diagnostic. Many other elements can also be detected, including C, Sr, and Ba. The Two Micron All Sky Survey (2MASS) IDs, right ascension (R.A.) and declination (Decl.), visual magnitudes ($V$), exposure times, Modified Julian Dates (MJD), slit widths, S/N ratios at 4000\,{\AA} and survey sources are listed in Table\,\ref{tab:ident}.

\subsection{Data Reduction and Radial-Velocity Measurements}

\begin{figure}[ht!]
\begin{center}
\vspace*{-0.4cm}
\hspace*{-0.9cm}
\includegraphics[scale=0.2]{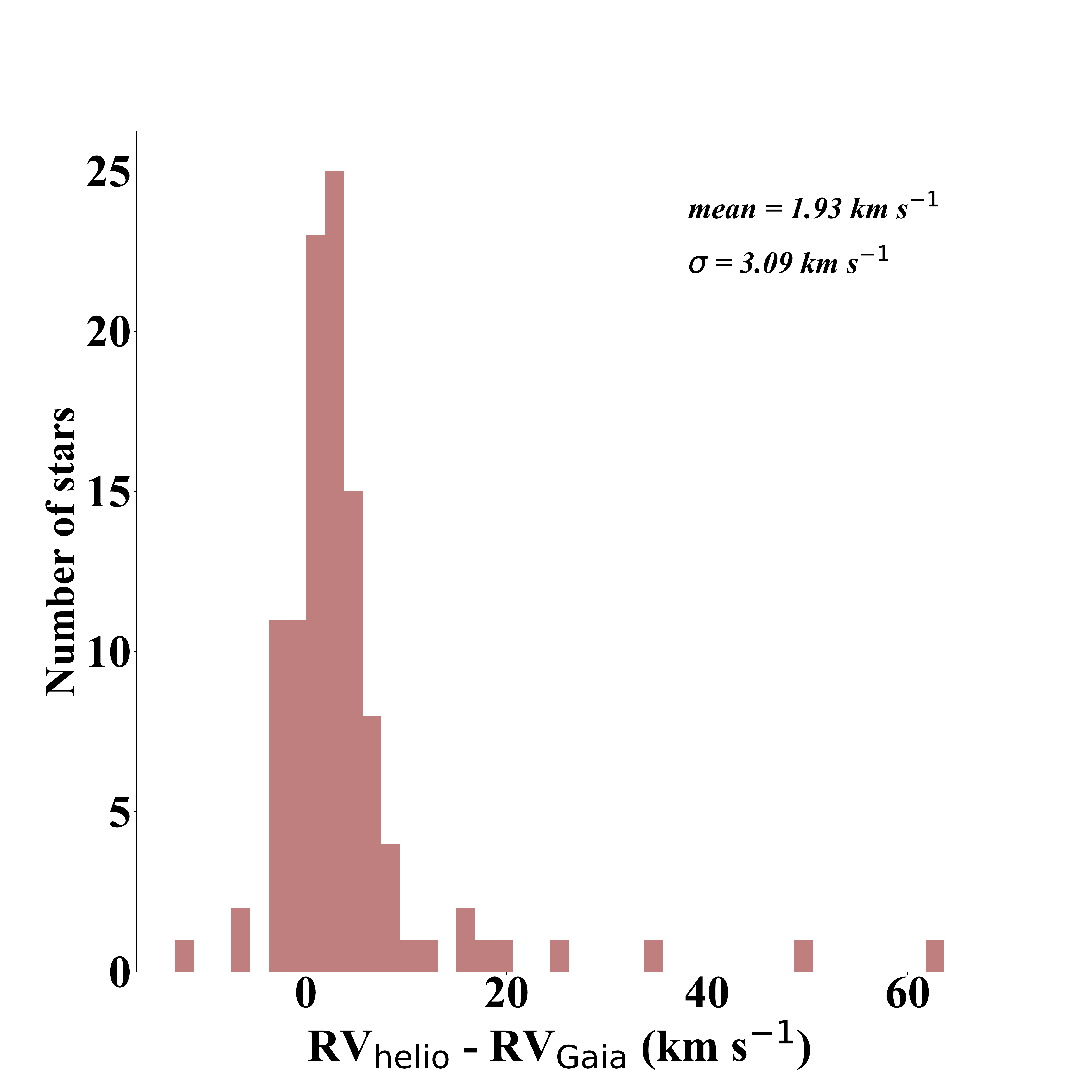}
\caption{\label{fig:rad_vel}Histogram showing radial-velocity differences of our targets, as measured in this study and by Gaia DR2. Eleven stars display significant differences $>1\sigma = 3.09$\,\kms\ (value determined excluding outliers). Seven of those stars were found to be SB2 binaries, and are labeled as such in Table\,\ref{tab:ident}.}
\end{center}
\end{figure}

Data were reduced using the latest versions of the  \texttt{Carnegie Python Distribution}\footnote{https://code.obs.carnegiescience.edu/mike} \citep{kelson2003}. Each order of each spectrum was then normalized and merged into a final spectrum, which was also 
radial-velocity shifted by cross-correlation with the \ion{Ca}{ii} line near 8542\,{\AA}
of the spectrum of HD\,140283, using the \texttt{Spectroscopy Made Hard} (\texttt{SMH}) analysis software (first described in \citealt{casey2014}). Heliocentric velocity (RV$_{\mathrm{helio}}$) corrections were then determined with the \texttt{rvcorrect} package in \texttt{IRAF}. We compare our RV$_{\mathrm{helio}}$ values to those from the $Gaia$ second data release (DR2; \citealp{gaia2018}), whenever available (110 stars). We present the comparisons in Figure\,\ref{fig:rad_vel}.
The radial velocities for eleven of the stars differ significantly from the $Gaia$ velocities, up to 60\,\kms. Excluding these outliers, our velocities agree with those from {\it Gaia} within a mean difference of 1.93\,\kms, with a standard deviation of $\sigma=3.10$\,\kms. We also compute the biweight-scale standard deviation\footnote{https://docs.astropy.org/en/stable/api/astropy.\\{\hspace*{0.55cm}stats.biweight.biweight\_scale.html}} of the full sample (including outliers). We find that $\sigma_{\mathrm{biweight}}=3.73$ agrees well with our reported $\sigma$.
Seven of those are found to be double-lined spectroscopic binaries (SB2), as their spectra exhibit clear contamination from a companion; these are labeled as such in Table\,\ref{tab:ident}. No spectral disentanglement of the two components was attempted for these stars, and thus they were not further analyzed with the rest of the sample.  We note that it is highly unlikely that the source of the \rproc\ enhancement is due to binarity, as it is unable to produce the neutron flux density ($n_{n}>10^{22}$\,cm$^{-3}$) needed to induce the \rproc\ \citep{thielemann2017}.

\section{Stellar Atmospheric Parameters}\label{sec:stell_param}

We determined stellar atmospheric parameters for the sample stars, namely the effective temperature (\teff), surface gravity (\logg), metallicity (\feh), and microturbulent velocity (\vt), from the abundances of \fei\ and \feii\ lines as determined from equivalent width (EW) measurements. The EWs of the lines were obtained by fitting Gaussian line profiles to the spectral absorption features using \texttt{SMH}. The \fei\ and \feii\ abundances were derived with 1D stellar atmospheric models, using three different spectroscopic methods: (i) assuming LTE, (ii) assuming LTE with \teff\ corrected to a photometric scale, following \citet{frebel2013}, as well as (iii) assuming NLTE. The atomic line-by-line properties for the \fei\ and \feii\ lines, including their EW and abundance measurements, are shown in Table\,\ref{tab:linelist}. The linelist is compiled from several data sources, and the $\log\,gf$ values for our lines are adopted from the up-to-date list compiled by \citet{roederer2018a} (see their Table\,2 and the references therein).

\subsection{LTE Stellar Parameters}\label{sec:lte}
The LTE stellar atmospheric parameters were estimated from the abundances of \fei\ and \feii\ lines using the 2017 version of the LTE radiative transfer code \texttt{MOOG} \citep{sneden1973}, including Rayleigh scattering treatment following \citet{sobeck2011}\footnote{https://github.com/alexji/moog17scat}.  Stellar 1D, LTE atmospheric models from \citet{castelli2004}, with standard $\alpha$-element enhancement of [$\alpha$/Fe] = +0.4 were used.
\teff\ values were determined by enforcing excitation equilibrium of the abundance of \fei\ lines as a function of excitation potential, $\chi$. The \logg\ values were determined by enforcing ionization equilibrium between the abundances inferred from the \fei\ and \feii\ lines. The \vt\ parameter was determined by ensuring no \fei\ abundance trends with reduced equivalent widths existed (REW; $\log(\mathrm{EW}/\lambda)$). The \feh\ abundances were determined from the average abundances of the \fei\ and \feii\ lines.
The number of \fei\ and \feii\ lines used, as well as the LTE stellar parameters derived for each star, are listed in Table\,\ref{tab:stell_param}. 
Stellar  parameter  uncertainties  are  determined assuming systematic uncertainties following the analysis in \citet{ji2016b}, and they are estimated to be up to 150\,K, 0.3\,dex, and 0.2\,\kms\ for \teff, \logg, and \vt, respectively. The \feh\ uncertainties are determined from the standard deviations of the \fei\ and \feii\ abundances.

\subsection{Corrected LTE Stellar Parameter Estimates}\label{sec:lte_corr}
Spectroscopic stellar parameters derived using the assumption of LTE have long been known to suffer from systematic uncertainties, especially for cooler metal-poor stars where atomic collisions are not sufficiently frequent to ensure that actual excitation and ionization equilibrium of the atomic populations is achieved. 
Determinations of \teff\ via photometric methods have been shown to be less prone to uncertainties compared to the LTE spectroscopic approach \citep{casagrande2010}. However, these require known absolute magnitudes and reddening toward the stars.
As a middle ground, we thus also determined the \teff\ of our target stars by ``correcting'' the LTE spectroscopic temperatures to a photometric scale, following the empirical calibration introduced in \citet{frebel2013}, hereafter denoted as \teff(FR13 corr.):

\begin{displaymath}
    {T_{\mathrm{eff}}\text{(FR13 corr.)} = {0.9\times} {T_{\mathrm{eff}}\text{(LTE)}} + 670}
\end{displaymath}

\begin{figure*}[ht!]
\begin{center}
\hspace*{-1.3cm}
\includegraphics[scale=0.2]{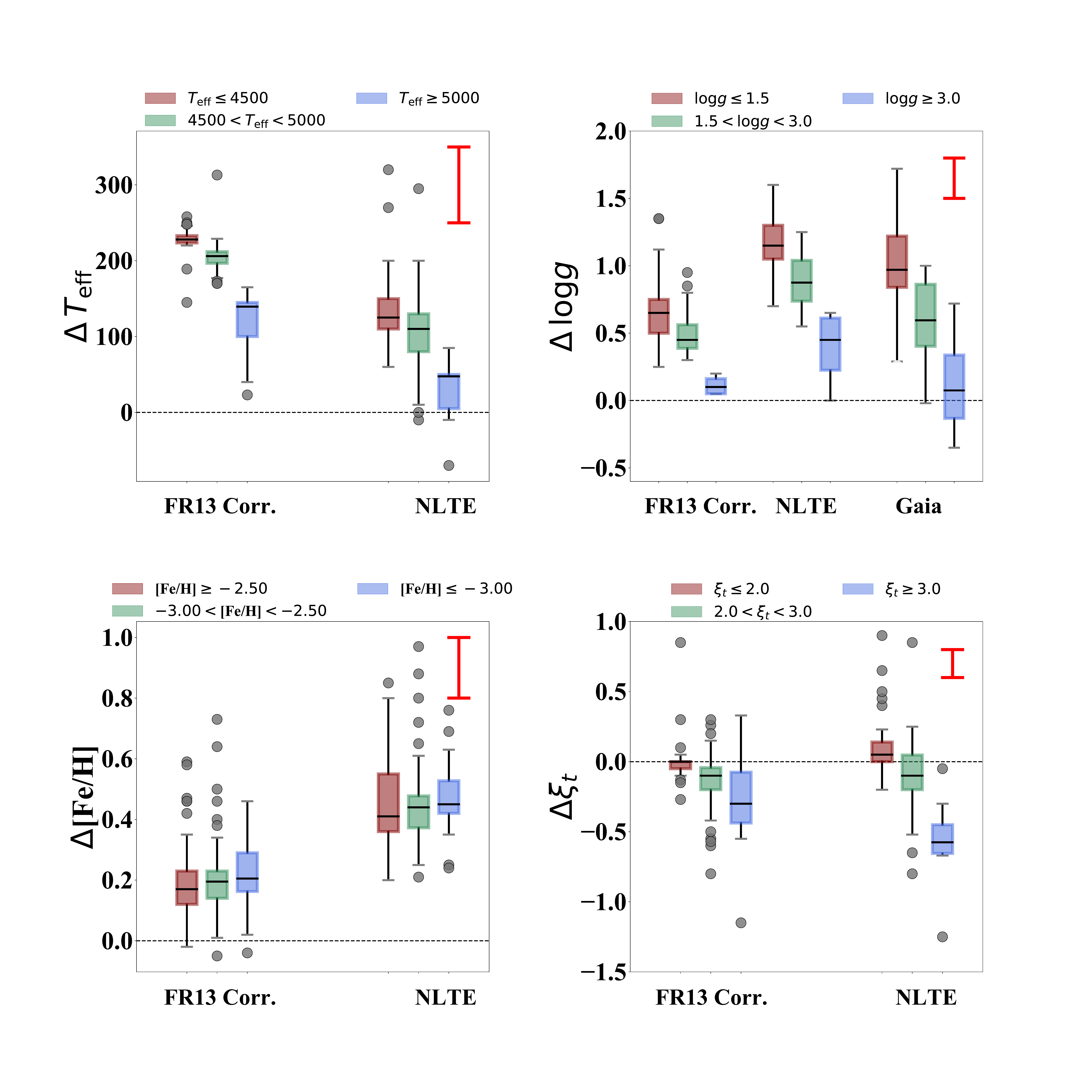}
\caption{\label{fig:nlte_corr}Boxplots showing the stellar parameter changes for \teff, \logg, \feh, and \vt, when comparing LTE results with those from applying the \citet{frebel2013} correction (labeled as FR13 Corr.), and from using a NLTE approach (labeled NLTE). 
The boxplots are color-coded for the ranges of stellar parameters (in LTE), as indicated on top of each panel. For \logg, we also show the differences arising from using $Gaia$ DR2 parallaxes and our uncorrected LTE values. Red uncertainty bars represent typical NLTE parameter uncertainties, corresponding to 100\,K for \teff, 0.3\,dex for \logg, 0.2\,dex for \feh, and 0.2\,\kms\ for \vt.}
\end{center}
\end{figure*}

For comparison to the ``strict'' LTE \teff\ results, we divide the stars into three categories covering different \teff\ ranges: \teff\ $\leq4500$\,K, $4500 < $ \teff\ $< 5000$\,K and \teff\ $\geq5000$\,K. We show the differences obtained relative to the strict LTE results in Figure \ref{fig:nlte_corr}, for each \teff\ range.
By design, the differences in the temperatures are more pronounced for cooler stars (\teff\ $<4500$\,K), by up to 250\,K, while smaller shifts were obtained for the hotter stars (\teff$>5000$\,K), up to 150\,K.

After fixing \teff\ to the FR13 corrected scale, we then also derived \logg, \vt, and \feh\ as explained in Section~\ref{sec:lte}, by forcing both ionization equilibrium and no abundance trends with line strengths. We compare the ``corrected'' \logg, \vt, and \feh\ to those determined strictly in LTE, as again shown in the boxplots in Figure\,\ref{fig:nlte_corr}. For \logg, the differences are presented for three surface gravity sub-categories: \logg\ $\leq1.5$, $1.5 < $ \logg\ $ < 3.0$ and \logg\ $\geq3.0$. The shifts from LTE are largest for the giant stars with \logg\ $\leq1.5$, up to $\sim +1.0$\,dex, while slightly smaller differences, up to +0.7\,dex, were obtained for the giants and sub-giants with $1.5 <$ \logg\ $< 3.0$. For stars with \logg\ $\geq 3.0$, the resulting differences were smaller still, up to +0.2\,dex, and lie within the LTE \logg\ uncertainties.

We also note that the differences obtained for \feh\ increase with decreasing metallicities, and are the largest for the extremely metal-poor stars (EMP) with \feh\ $\leq-3.0$ (Figure\,\ref{fig:nlte_corr}), reaching up to +0.5\,dex. On average, the corrections are $\sim +0.2$\,dex.  The corrections for \vt\ are negative relative to the uncorrected LTE case, and are decreasing with increasing \vt, reaching $-0.5$\,\kms\ for \vt\  $\geq3.0$\,\kms.

\begin{figure*}
\begin{center}
\vspace*{-0.4cm}
\hspace*{-1cm}
\includegraphics[scale=0.2]{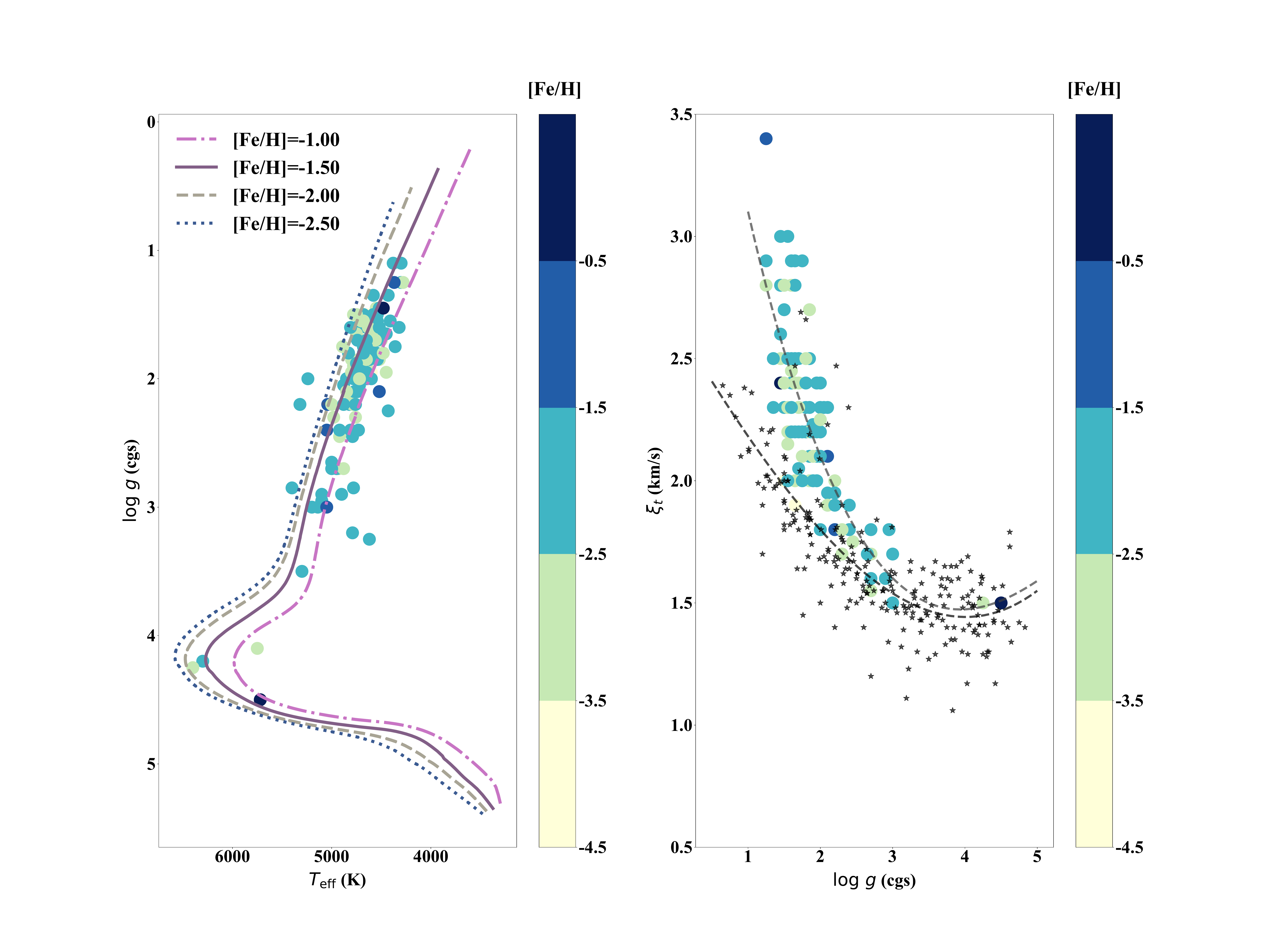}
\caption{\label{fig:isoch}Left panel: NLTE stellar parameters of the target stars (\teff\ and \logg\ color-coded as a function of \feh), over-plotted with Dartmouth isochrones at 12.5\,Gyrs for four different metallicities, $\feh\ =-1.00$, $\feh\ =-1.50$, $\feh\ =-2.00$, and $\feh\ =-2.50$, respectively \citep{dartmouth2008}. Right panel: NLTE \logg\ versus \vt\ for the target stars, color-coded as a function of \feh. The polynomial fit to the data is shown by the gray dotted line. Also shown, for comparison, are the stars (show by the small star markers) from \citet{cayrel2004} and \citet{barklem2005}, as well as the polynomial fit to their \logg-\vt\ (black dotted line).}
\end{center}
\end{figure*}

\subsection{NLTE Stellar Parameter Estimates}\label{sec:nlte}

Atmospheric stellar parameter determinations for metal-poor stars from LTE spectroscopic methods are affected by unaccounted-for departures from statistical equilibrium that can introduce significant systematic uncertainties, since line formation and populations of non-dominant species (in this case \fei) can potentially deviate from the Saha–Boltzmann equilibrium assumed in LTE \citep{lind2012,amarsi2016,Ezzeddine2017}. To account for such departures, especially in giant and warm stars, it is necessary to investigate the formation of iron lines  in NLTE.
Therefore, we also determine stellar parameters for our target stars using 1D, NLTE radiative transfer models.

\subsubsection{NLTE Methods}

NLTE abundances were computed for \fei\ and \feii\ lines from their EWs using the radiative transfer code \texttt{MULTI} in its 2.3 version \citep{Carlsson1986,carlsson1992}, and employing 1D \texttt{MARCS} model atmospheres \citep{gustafsson1975,gustafsson2008} interpolated to the corresponding parameters. Blanketing from background opacities, excluding Fe lines, were employed from the \texttt{MARCS} opacity package (B. Plez, private communication).

The \fei/\feii\ atomic model used in the NLTE calculations is described in more detail in \citet{Ezzeddine2017}. This model was built adopting up-to-date atomic data. Quantum computations of inelastic collision rates of \fei\ with neutral hydrogen atoms were implemented from \citet{barklem2018}. These collisions play an important dominant role (over electrons) in NLTE calculations of cool stars.

The NLTE stellar parameters were derived following the procedure in \citet{Ezzeddine2017}, where first approximations of the parameters were derived simultaneously by multi-dimensional fitting of the observed \fei\ and \feii\ EW of the target stars to theoretical NLTE EW in a pre-computed dense grid of \teff, \logg, \vt, and \feh\ parameter space, using the Levenberg-Marquardt algorithm. The excitation and ionization equilibrium, as well as the trends with reduced EW, were then checked for each star, ensuring no trends.

\subsubsection{NLTE Results}
The derived NLTE stellar parameters are shown in Table\,\ref{tab:stell_param}. For \teff, \logg, and \vt, we adopt typical total uncertainties arising from the uncertainties in the atomic data and abundance slopes versus $\chi$ and REW, of 100\,K in \teff, 0.3\,dex in \logg, and 0.2\,\kms\ in \vt. \feh\ uncertainties were adopted from the standard deviations of the line-by-line abundances and are shown in  Table\,\ref{tab:stell_param}, which vary from 0.08 to 0.18\,dex.  
The NLTE gravities, as a function of \teff\ and \feh, are shown on the left panel of Figure~\ref{fig:isoch}, over-plotted on four 12.5\,Gyr Dartmouth isochrones \citep{dartmouth2008}, with [$\alpha$/Fe] = +0.4 and \feh\ $=-1.00$, \feh\ $=-1.50$, \feh\ =$-2.00$, and \feh\ $=-2.50$, respectively. 
 The stars mostly occupy the upper portion of the giant branch in the \teff-\logg\ diagram. This likely stems from our target selection bias toward cool bright giants for the search for $r$-process stars by the RPA \citep{placco2018,placco2019}. We note that, despite this bias, the RPA sample in this work, as well as other RPA data releases, are expected to be fairly representative, as all initial “snapshot” spectra of the target stars were exposed to determine a reliable r-process enhancement measurement, by adjusting exposure times accordingly to obtain
S/N ratios $> 30$ (per pixel) in the blue around the \euii\ line at 4129\,{\AA} (a representative line of \rproc\ enhancement, see Section\,\ref{sec:rproc_abund} below).
Four of the stars are found to be hot dwarfs (\teff\ $>5500$\,K, \logg\ $>4.0$), while another two stars are field horizontal-branch stars.  Additionally, one of the stars, 2MASS~J09291557+0838002, is a newly identified ultra-metal-poor (UMP) star at \feh(NLTE)$=-4.02$. Four \rproc-enhanced stars at the \rI, \rs, and \limr\ levels are detected at \feh\ $>-1.5$. We also plot, on the right panel of Figure\,\ref{fig:isoch}, \logg\ versus \vt\ for our target stars. For comparison, we also show the results from previous studies by \citet{cayrel2004} and \citet{barklem2005} for their very metal-poor star samples. We fit the \logg-\vt\ dependence for both our target stars, as well as those from previous studies with third degree polynomials\footnote{The fitting coefficients for the third degree polynomial,  y = ax$^3$ + bx$^2$ + cx + d (where x and y correspond to \logg\ and \vt, respectively), obtained for our target stars are a = $-0.02$, b = 0.37, c= $-1.98$, and d = 4.73.}.
We find that \vt\ increases for giant stars with lower \logg\ as compared to the dwarfs. A similar trend is obtained from previous studies, as shown by the black star markers and their third degree polynomial fit on the same Figure, although the NLTE dependence from our study shows a more severe increase for \vt\ for the giant stars. These positive NLTE \vt\ corrections are nominally driven by the abundance differences obtained between LTE and NLTE from the stronger saturated \fei\ lines (with $EW>50$\,m{\AA}), as explained in the detailed NLTE study by \citet{lind2012}. 

We also compare the NLTE stellar parameters from our sample and from the RPA-1 and RPA-2 papers \citep{rpa1,rpa2} to their corresponding LTE values (Figure\,\ref{fig:nlte_corr}).
We find that the NLTE temperatures are higher than those determined in LTE, and the shifts are more pronounced for cooler stars, reaching up to +200\,K. On average, the \teff\ shifts between NLTE and LTE are +150\,K for \teff\ $\leq4500$\,K, +120\,K for $4500 <$ \teff\ $\leq5000$\,K and +50\,K for \teff\ $>5000$\,K. The NLTE corrections (\teff(NLTE)$-$ \teff(LTE)) are, however, lower than those obtained by using the FR13 corrections, by $\sim 100$\,K, on average.

The NLTE corrections in \logg, defined by \logg(NLTE) $-$ \logg(LTE), are also positive, and found to be increasing toward lower gravities, up to +1.2\,dex. These corrections agree with those derived for the  stars in \citet{rpa2}. On average, the corrections are +1.2\,dex for \logg\ $\,\leq1.5$, +1.0\,dex for $1.5 <$ \logg\,$ < 3.0$, and +0.5\,dex for \logg\ $\geq\,$ 3.0. The corrections in \feh\ are also increasing toward lower metallicities, in line with other NLTE predictions \citep{mashonkina2011,lind2012,amarsi2016}, and can reach up to +0.6 for \feh\ $\leq-3.0$, +0.45 for $-3.0 < $ \feh\ $ <-2.0$, and +0.3 for \feh\ $\geq-2.0$.
We note that these metallicity NLTE corrections are expected to be larger, on average, than those nominally obtained by fixing the \teff\ and \logg\ values prior to determining the LTE and NLTE \feh\ (e.g., \citealt{lind2012,mashonkina2011}). Therefore, the \feh\ corrections obtained in our study also reflect the abundance changes due to the \teff\ and \logg\ NLTE corrections obtained for the target stars.
Finally, the NLTE \vt\ corrections are negative, and agree on average with those obtained by applying the FR13 corrections. The corrections are, however, slightly lower in NLTE for \vt\ $\geq$ 3.0\,\kms, and can reach $-0.6$\,\kms.

We note that any differences in the atmospheric models used in the NLTE and LTE calculations (\texttt{MARCS} vs. \texttt{ATLAS}) are expected to have negligible effects on the resulting stellar parameters in LTE. \citet{roederer2014b} showed in their analysis of a large number of metal-poor stars, including both giants and subgiants, that differences obtained between both models are less than 8\,K for \teff, 0.005\,dex for \logg, 0.002\,\kms\ for \vt, and 0.006\,dex for \feh. These values are negligible as compared to the differences obtained for the stellar parameters between NLTE and LTE.

\subsection{Estimates of $log\,g$ from $Gaia$ DR2 Parallaxes}\label{sec:gaia}
For \logg, we also derive surface gravities from $Gaia$ DR2 parallax-based distances \citep{bailer-jones2018}. To calculate distance-based surface gravities, we adopt a relation between \logg,
stellar mass, $V_{\mathrm{mag}}$, and bolometric corrections.
For the dwarfs and giants in our sample, a mass of 0.7 and 0.8 solar masses was adopted, respectively. The RAVE $V_{\mathrm{mag}}$ were used, and the bolometric corrections were adopted from \citet{casagrande2014}. 

The derived \logg$^{Gaia}$ values are shown in the last column of Table~\ref{tab:stell_param}. For comparison, we also show the differences  between \logg$^{Gaia}$ and those derived in LTE in Figure~\ref{fig:nlte_corr}. The gravity corrections are more pronounced for the giant stars with \logg\ $\leq1.5$, with an average shift of $\sim +1.2$\,dex. Slightly lower shifts are obtained for stars with $1.5 < $ \logg\,$ < 3.0$, of up to +0.6\,dex, and for stars with \logg\ $\geq\,$ 3.0, up to +0.1\,dex. The $Gaia$ gravities agree with those obtained in NLTE within uncertainties of $\sim$0.3\,dex, on average.

We note that 26 of our sample stars have  {\it Gaia} DR2 parallax-based distances larger than 6\,kpc \citep{bailer-jones2018}. Those stars are likely to lead to larger uncertainties in the derived \logg$^{Gaia}$ ($>0.2$\,dex), due to the larger systematic shifts of their {\it Gaia} DR2 parallaxes \citep{arenou2018}. We flag these stars as such in Table\,\ref{tab:stell_param}, and warn against the use of their {\it Gaia} \logg.

Given that the NLTE \logg\ values agree overall with those from non-spectroscopic ones ($Gaia$), and that the NLTE stellar parameters are considered to be better representatives of the stellar atmospheres of metal-poor stars than LTE (e.g., see review by \citealt{asplund2005}), we decided to adopt the NLTE stellar parameters for our target stars, and use them to determine the chemical abundances of our stars.
Nevertheless, our analysis clearly shows that in the absence of a full NLTE approach, the FRE13 values (calibrated to photometric \teff) provide reasonable stellar parameters, lying within the uncertainty levels of the corresponding NLTE values, where typical values of 100\,K for \teff, 0.3\,dex for \logg, 0.2\,dex for \feh, and 0.2\,\kms\ for \vt\ are shown by the red error bars on Figure\,\ref{fig:nlte_corr}.

\section{Chemical Abundances}\label{sec:abund}

We derive abundances for the light elements, $\alpha$ elements, and Fe-peak elements for the target stars, including C, O, Na, Mg, Al, Si, K, Ca, Sc, \tii, \tiii, V, Cr, Mn, Co, Ni, and Zn using \texttt{MOOG}. Additionally, we derive the abundances for the neutron-capture elements Eu, Ba, and Sr, in order to sort our target stars into the sub-classifications suggested by the reviews of \citet{Beers2005} and \citet{frebel-rev2018}. The abundance ratios [X/H] are calculated relative to the solar abundances presented by \citet{asplund2009}. The abundances (except for C, O, Al, and the neutron-capture elements) were derived from a line-by-line analysis using EWs of the lines and a curve-of-growth (COG) method. Sufficiently weak lines lying on the linear part of the COG were chosen with REW $\leq -4.5$. The remaining abundances were derived using spectrum synthesis of their corresponding lines, taking into account isotopic ratios.  Linelists for each region of interest are generated with \texttt{linemake}\footnote{https://github.com/vmplacco/linemake} \citep{sneden2008}.  The Hyper-fine structure (HFS) was considered for the Fe-peak elements, including Sc, V, Mn, and Co, when necessary.
For carbon, an average isotopic shift ratio of  $^{12}$C/$^{13}$C = 4/1 was used. This was determined for each star individually, by fitting the lines around the 4313\,{\AA} CH feature, specifically, those at 4217\,{\AA}, 4225\,{\AA}, as well as 4302\,{\AA}. Upper limits were determined by matching the noise levels in the observed spectral lines with the corresponding synthetic spectral lines. Our line-by-line atomic data, EW and abundances are listed in Table\,\ref{tab:linelist}. The average abundances are also shown in Table\,\ref{tab:abund_fe_peak}.

\subsection{Neutron-Capture Element Abundances}\label{sec:rproc_abund}
\begin{figure*}[ht!]
\begin{center}
\hspace*{-1.0cm}
\includegraphics[scale=0.2]{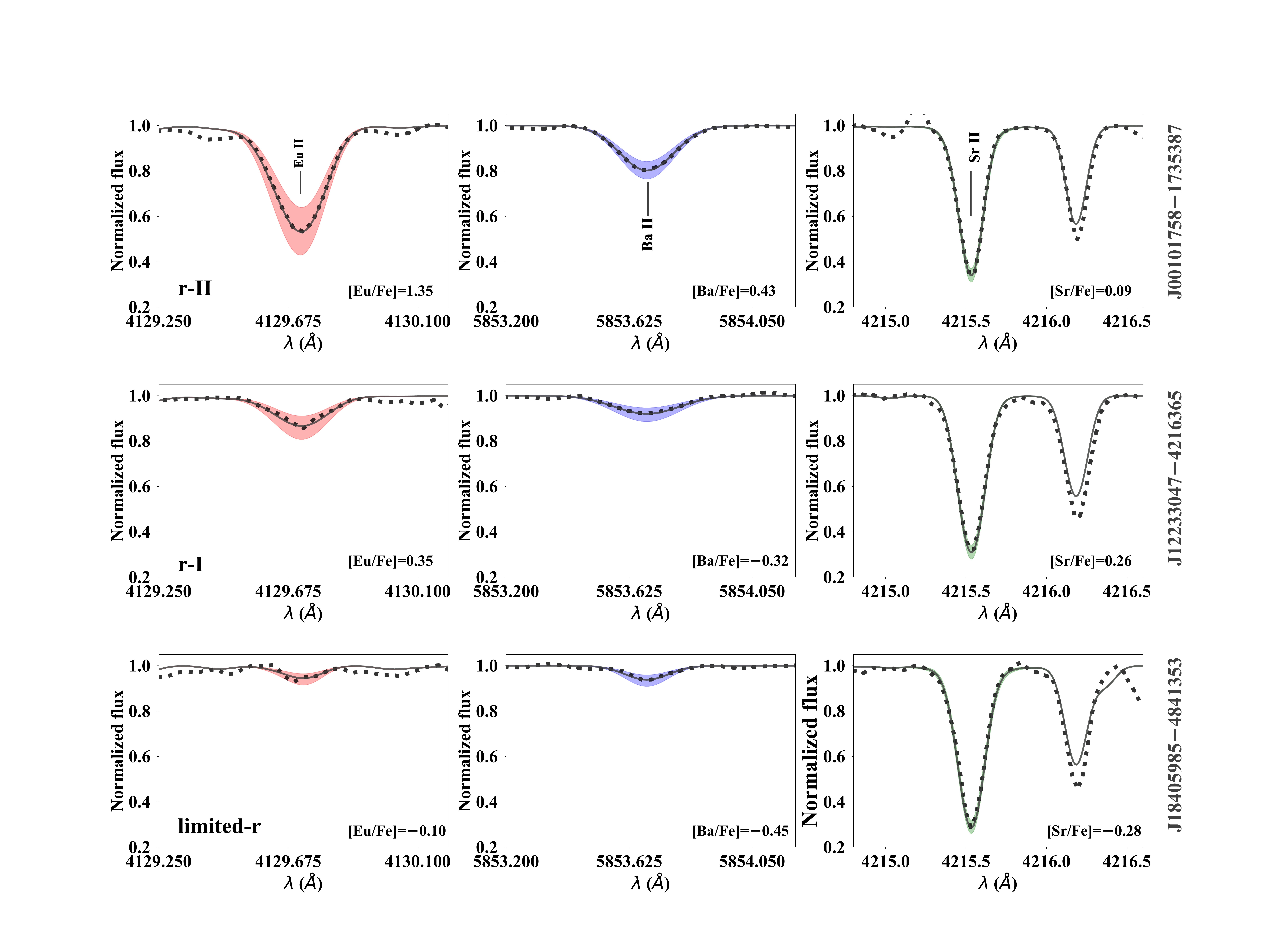}
\caption{\label{fig:obs_spec}Observed spectral lines (dotted lines) and their abundance syntheses (gray lines) of \srii\ at 4215\,{\AA}, \baii\ at 5853\,{\AA},  and \euii\ at 4129\,{\AA} for \rII\ (J00101758$-$1735387, top panel), \rI\ (J12233047$-$4216365, middle panel), and \limr\ (J18405985$-$4841353, bottom panel) stars discovered in our sample. Representative uncertainties of $\pm 0.2$\,dex are shown for each lines by the filled syntheses colors.}
\end{center}
\end{figure*}

Strontium abundances were derived from the \srii\ lines at 4077\,{\AA}, 4161\,{\AA}, and 4215\,{\AA}. With the line at 4077\,{\AA} being saturated in most of the stars, and the line 4161\,{\AA} being very weak, especially in EMP stars, the Sr abundances were primarily derived from the line at 4215\,{\AA}.  We note that the strong 4077 and 4215\,{\AA}
\srii\ lines may be affected by NLTE effects, however, \citet{short2006} reported an offset of $<0.1$\,dex for these lines for stellar parameters similar to those of our sample of stars. Barium abundances were derived mainly from the two \baii\ lines at 5853\,{\AA} and 6141\,{\AA}, which are not saturated, unlike the line at 4554\,{\AA}. For Eu, the abundances were derived from several \euii\ lines at 3907\,{\AA}, 4129\,{\AA}, 4205\,{\AA}, 4435\,{\AA}, 6437\,{\AA}, and 6645\,{\AA}, following \citet{rpa1}. The bluer line at 3907\,{\AA} was used when the blue S/N of the spectra was high enough ($\geq80$). However, the strongest Eu line, located at 4129\,{\AA}, was was the primary line used for most stars. Examples of observed Sr, Ba, and Eu lines at 4215\,{\AA}, 5853\,{\AA}, and 4129\,{\AA}, respectively, for representative \rII, \rI, and \limr\ stars, as well as their abundance syntheses, are shown in Figure\,\ref{fig:obs_spec}.

The abundance ratios [Sr/Fe], [Ba/Fe] and [Eu/Fe], calculated with NLTE \feh\ values, are listed in Table\,\ref{tab:rproc_clas}. They 
were also used to classify our stars into their sub-classes following \citet{frebel-rev2018}:

\begin{itemize}
    \item \rII:  [Eu/Fe] $ > +1.0$ and [Ba/Eu] $ < 0.0$
    \item \rI: $+0.3 \leq $ [Eu/Fe] $ \leq +1.0$ and [Ba/Eu] $ < 0.0$
    \item \limr: [Eu/Fe] $ < +0.3$, [Sr/Ba] $ > +0.5$, and [Sr/Eu] $ > 0.0$
    \item $s$-process: [Ba/Fe] $ > +1.0$ and [Ba/Eu] $ > +0.5$
    \item $r/s$: $0.0 < $ [Ba/Eu] $ < +0.5$
    \item CEMP-$r$: [C/Fe] $> +0.7$, [Eu/Fe] $ > +0.3$, and [Ba/Eu]$ < 0.0$
    \item CEMP-no: [C/Fe] $ >+0.7$ and [Ba/Fe] $ < 0.0$
    \item Non-\rproc\ enhanced (non-RPE):  [Eu/Fe] $ < +0.3$, [Sr/Ba] $ \leq 0.5$, and [Sr/Eu] $ < 0.0$
\end{itemize}

\begin{figure*}[ht!]
\begin{center}
\hspace*{-1.3cm}
\includegraphics[scale=0.42]{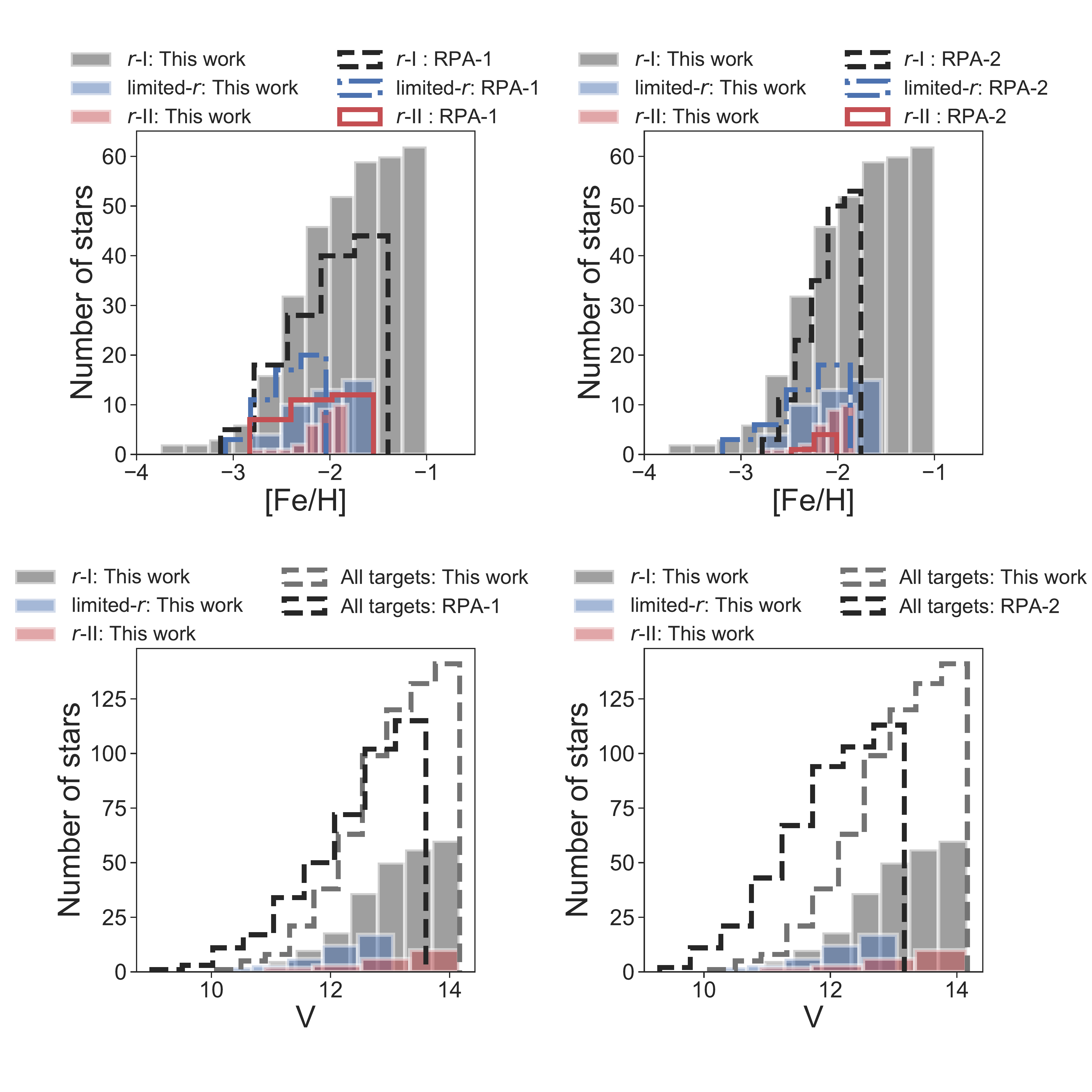}
\caption{\label{fig:hist}Cumulative histograms showing the number of \rI\ (filled gray), \rII\ (filled red), and \limr\ (filled blue) stars discovered in this work, as a function of NLTE \feh (upper panels) and V magnitudes (lower panels). Also shown, for comparison, are all the stars discovered and analyzed in the RPA-1 \citep{rpa1} and RPA-2 \citep{rpa2} samples, as well as the total targets (dashed gray histogram) analyzed in this work (including non-RPE).}
\end{center}
\end{figure*}

\begin{figure}[ht!]
\begin{center}
\hspace*{-1.0cm}
\includegraphics[scale=0.42]{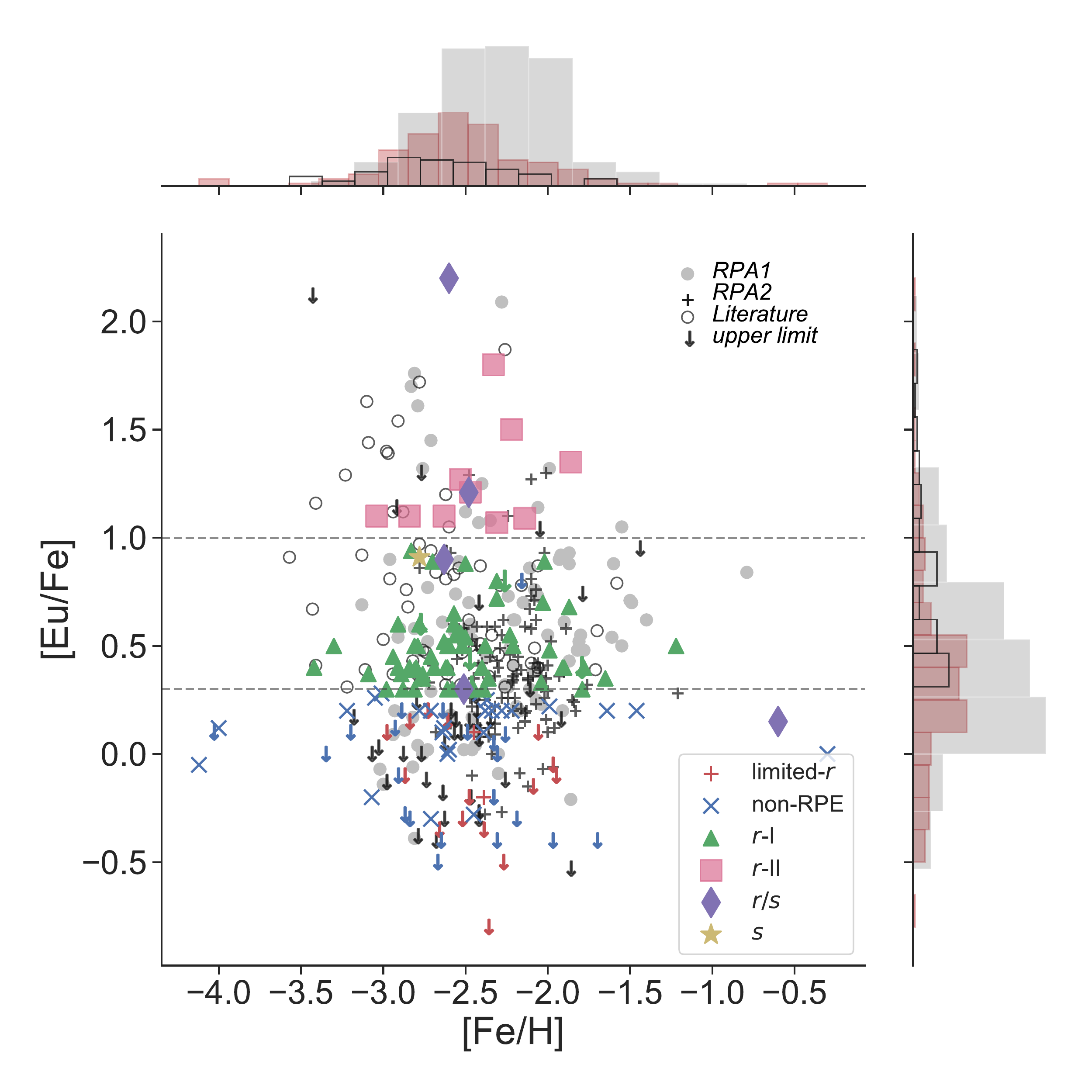}
\caption{\label{fig:abund_eufeh} Derived [Eu/Fe] abundances, as a function of [Fe/H], for the \rI, \rII, \limr, \rs, as well as non-RPE stars. Also shown are the stars from the RPA-1 \citep{rpa1} and RPA-2 \citep{rpa2} samples, as well as \rI\ and \rII\ stars from the literature (extracted from JINAbase \citealt{jinabase2018}). 
The arrows display upper limit measurements. Statistics of the RPA-3 stars are shown by marginal red histograms for both the [Eu/Fe] and [Fe/H] ranges, and those of RPA-1 and RPA-2 stars are shown by marginal gray histograms. \rI\ and \rII\ stars from the  literature are shown by open black histograms. Literature values are taken from \citet{westin2000}, \citet{johnson2002}, \citet{cayrel2004}, \citet{honda2004}, \citet{barklem2005}, \citet{ivans2006}, \citet{mashonkina2010}, \citet{SCM2014}, and \citet{roederer2014b}. Upper limits from the literature data have been suppressed for clarity.}
\end{center}
\end{figure}

\begin{figure}[ht!]
\begin{center}
\hspace*{-0.5cm}
\includegraphics[scale=0.42]{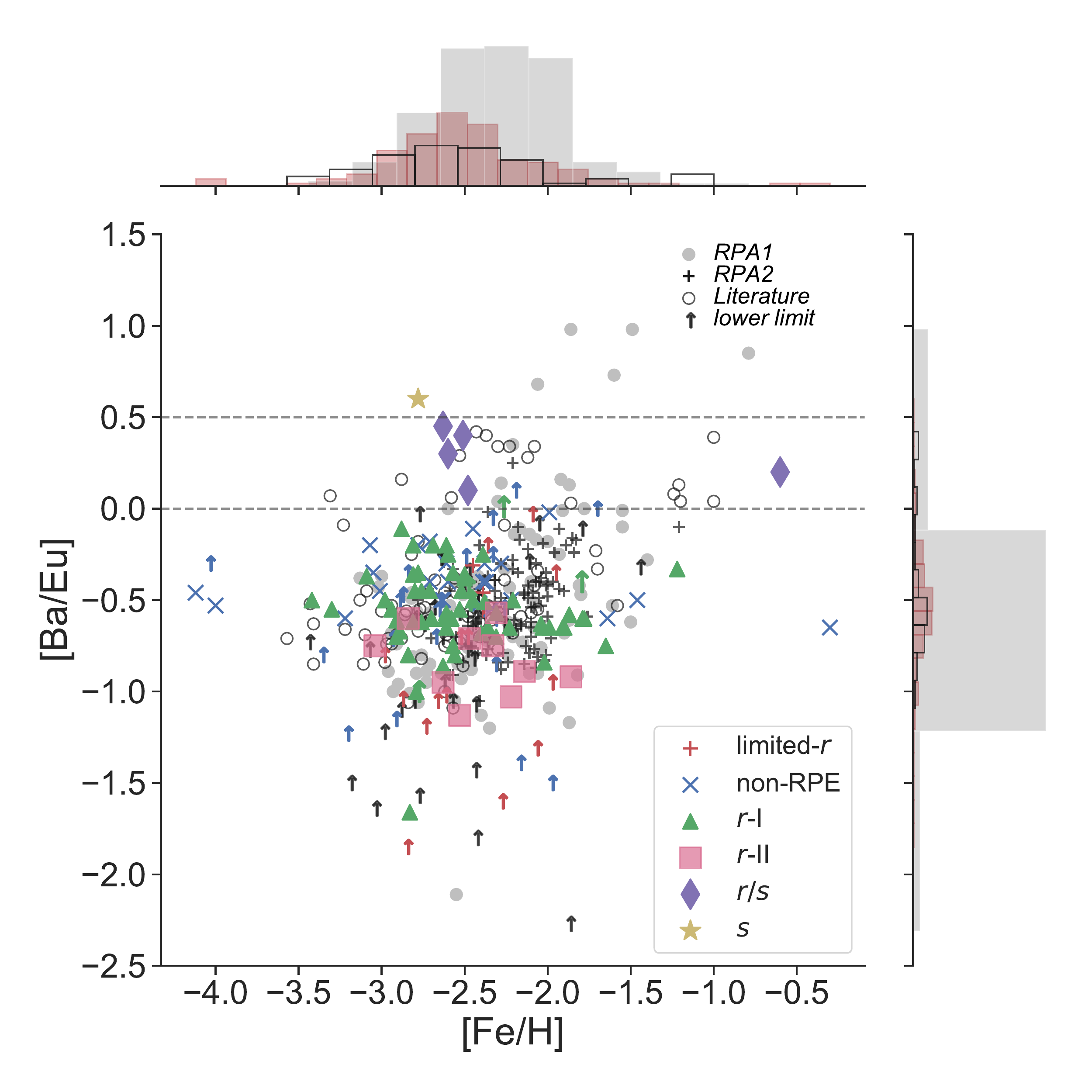}
\caption{\label{fig:abund_baeu} Same as in Figure~\ref{fig:abund_eufeh}, but for [Ba/Eu] abundance ratios. Dividing dotted lines indicating differences between the $s$-process-enhanced stars ([Ba/Eu] $ > +$0.5), \rs-enhanced stars ($0.0 < $ [Ba/Eu] $ <$ +0.5), and \rproc-enhanced stars ([Ba/Eu] $ < $ 0.0) abundance ratios are also shown.
}
\end{center}
\end{figure}

\begin{figure}[ht!]
\begin{center}
\hspace*{-1.0cm}
\includegraphics[scale=0.42]{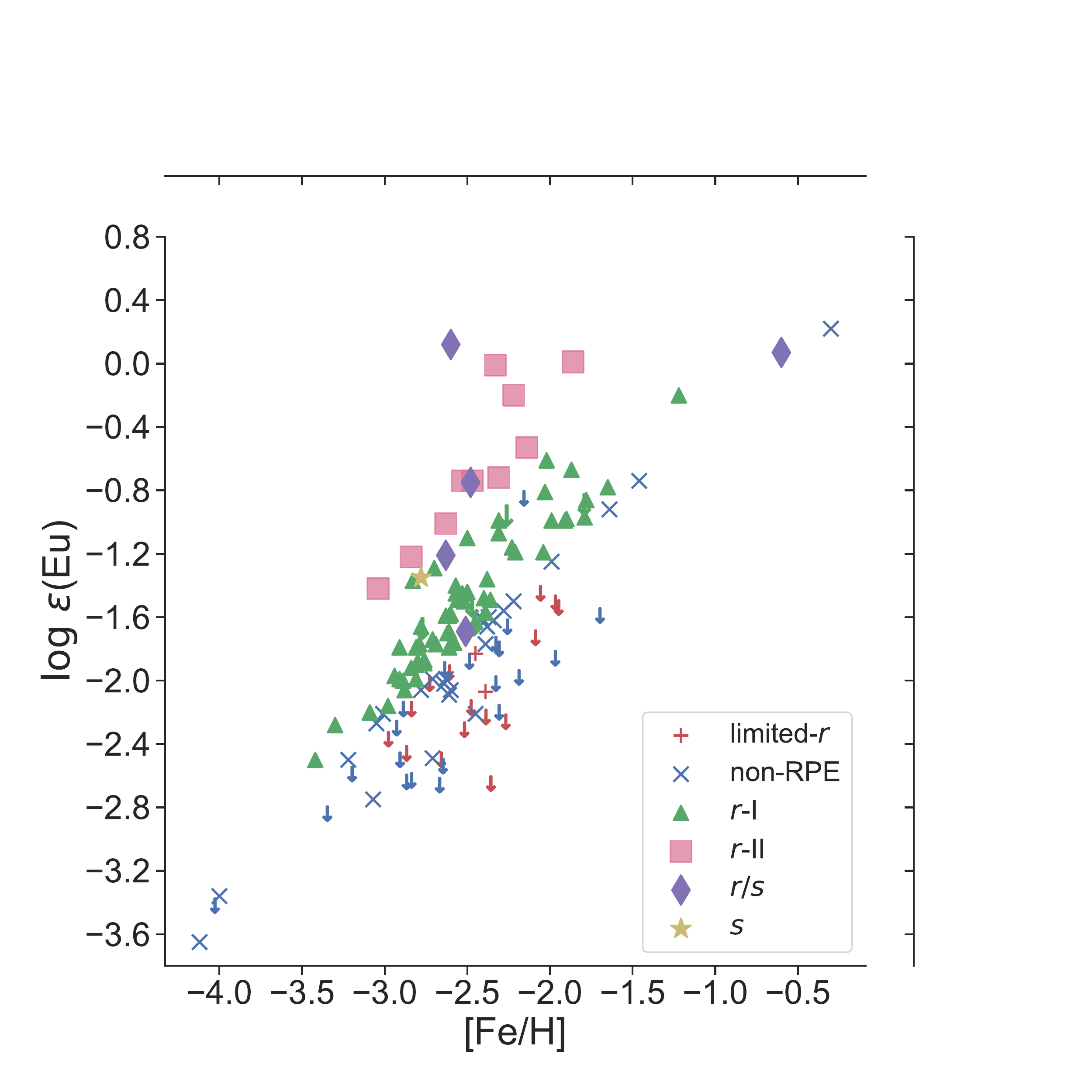}
\caption{\label{fig:eu_abs} Absolute Eu abundances, as a function of [Fe/H], for the \rII, \rI, \limr, \rs, and $s$-process-enhanced stars identified in this work.}
\end{center}
\end{figure}

\begin{figure*}[ht!]
\begin{center}
\hspace*{-2.0cm}
\includegraphics[scale=0.225]{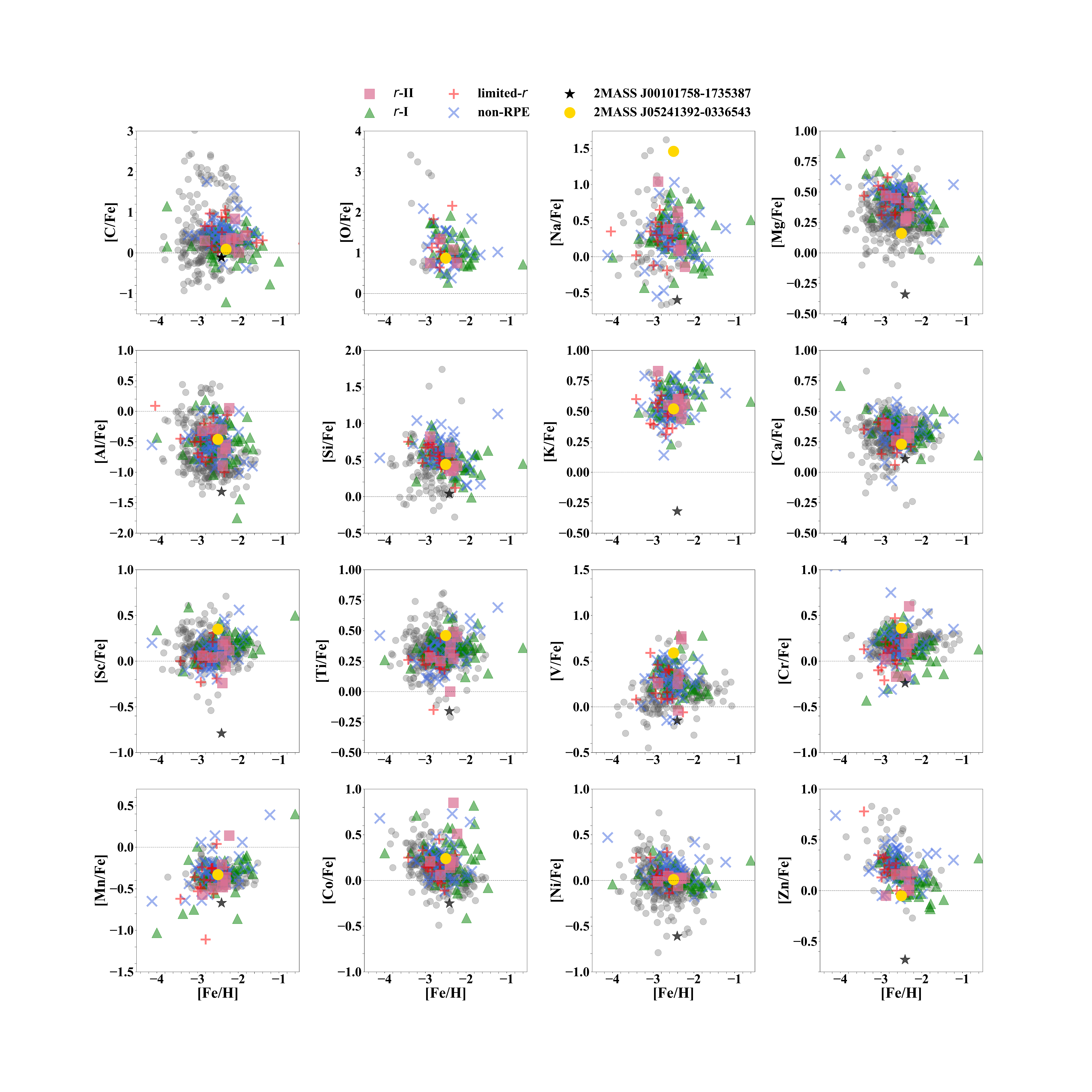}
\caption{\label{fig:abund_rel}Light, $\alpha$ and fe-peak abundance ratios (relative to Fe) of the target stars, as a function of [Fe/H], for the different \rproc\ sub-classifications. Literature studies (shown as gray points) by \citet{cayrel2004}, \citet{barklem2005}, and \citet{cohen2013}, and \citet{roederer2014b} are also shown for comparison. Stars with distinguished abundance patterns (see discussion in Section\,\ref{sec:elem_abund}), 2MASS~J00101758$-$1735387 and 2MASS~J05241392$-$0336543, are also shown by the black star and yellow circle, respectively.}
\end{center}
\end{figure*}

\subsection{CO and the Light-Element Abundances}\label{sec:light-el_abund}

We derive the C abundances via spectrum synthesis of the molecular CH $G$-band at 4313\,{\AA} \citep{masseron2014}. As most of our stars lie on the red giant branch (see Figure\,\ref{fig:isoch}), the carbon abundances are assumed to be depleted by the CN cycle having operated during the stars' evolution. Relative [C/Fe] ratios of the target stars were therefore corrected for evolutionary status following \citet{placco2014}. We show both uncorrected and corrected values in Table\,\ref{tab:rproc_clas}. Additionally, we derive abundances of the light elements O and Na, using the forbidden \oi\ lines at 6300 and 6363\,{\AA} for O (whenever detectable), and the \nai\ doublet at 5889 and 5895\,{\AA} for Na, if the lines are sufficiently weak. We note that \citet{lind2011} reported non-significant NLTE abundance corrections of $<0.1$\,dex for the \nai\ doublet at our metallicity range. The [O/H] and [Na/H] abundances are listed in Table\,\ref{tab:abund_light}.

\subsection{The $\alpha$-element, K, and Al abundances}\label{sec:alpha-el_abund}
We derive $\alpha$-element abundances for \mgi, \cai, \tii, and \tiii\ from the EWs of their corresponding lines.  Potassium abundances were determined when the \ki\ lines at 7664 and 7698\,{\AA} were found to be unpolluted by telluric lines.
For Al, we determine the abundances by synthesizing the blue resonance \ali\ line at 3961\,{\AA}, taking into account the C abundance.  We caution that this line can be prone to significant 3D and NLTE corrections, up to 0.6\,dex for our range of metallicities, as reported by \citet{nordlander2017}.

\subsection{Iron-Peak Element Abundances}\label{sec:fepeak_abund}
We also derive the Fe-peak abundances for our target stars using \scii, \vii, \crii, \mni, \coi, \nii\ and \zni\ from their EWs using a multitude of lines, whenever available. We did not derive abundances from \vi\ and \cri\ lines, as they are prone to larger NLTE effects than their corresponding ionized species, \vii\ and \crii\ \citep{bergemann2010}.  We note that recent results by \citet{bergemann2019} showed that \mni\ lines can be prone to large NLTE and 3D effects, which they found to be increasingly larger for lower metallicities. For our sample of stars in this work, with [Fe/H] ranging typically between $-3.0<\mathrm{[Fe/H]}<-2.0$, the authors report 3D, NLTE corrections between 0.3 and 0.6\,dex for \mni\ lines. We do not include these corrections in this paper, but anticipate to include them in future papers once the correction grids have been made publicly available.

\section{Discussion}\label{sec:disc}

\subsection{$R$-Process-Enhanced Stars}
Following the neutron-capture sub-classifications defined in \citet{Beers2005} and \citet{frebel-rev2018}, we identify in this work 10 new \rII, 62 new \rI, and 17 new \limr\ stars. Additionally, we find one \sproc-enhanced and five \rs\ stars.
Overall, 46 stars from our target sample of 141 stars are non-\rproc-enhanced (hereafter non-RPE, following the definition in \citealt{rpa2}). This shows that our overall selection procedure works well. One of our non-RPE stars, 2MASS~J00452379$-$2112161, is a rediscovery of the ultra metal-poor star CD$-38\,^{\circ}245$ \citep{bessell1984}.
Seventeen of the 141 stars were found to be Carbon-enhanced metal-poor (CEMP) stars, one enhanced at the \rII\ level, four at the \rI\ level, and three are \limr\ stars. Additionally, three out of five of the \rs-enhanced stars are CEMP stars, as well as the \sproc-enhanced star. Five stars were also found to be CEMP-no stars.

This brings up the total number of $r$-process-enhanced stars (at the \rII, \rI, \limr, and \rs\ levels) identified by the RPA efforts from this work, \citet{rpa1}, \citet{rpa2}, \citet{roederer2018a}, \citet{holmbeck2018}, \citet{cain2018}, \citet{gull2018}, and \citet{placco2017} to 253 out of a total of 381 stars. Of those, a total of 28 stars are enhanced at the \rII\ level (7.4\%), 163 stars at the \rI\ level (42.7\%), 56 stars at the \limr\ level (14.7\%), and 6 stars are enhanced at the \rs\ level (1.6\%). We caution that these values should not be used as physically interpretable estimates of the percentage of r-process enhanced stars.  They likely represent upper bounds as some stars were selected for publication on account of their high level of r-process enhancement \citep{placco2017,cain2018,gull2018,holmbeck2018}, and other sample selection biases have not yet been fully quantified.  The RPA intends to report more definitive estimates of the percentage of r-process enhanced stars in future work.

Figure~\ref{fig:hist} shows the cumulative histograms of newly identified \rI, \rII, and \limr\ stars in this work, as a function of metallicity (upper panels) and V magnitudes (lower panels). None of those were previously identified in the RPA-1 and RPA-2 samples (open histograms; \citealp{rpa1,rpa2}).
Similar to the results found in the RPA-1 and RPA-2 samples (also shown on the plots), we find  \rI\ and \rII\ stars over a wide metallicity range, from \feh\ = $-3.5$ to \feh\  = $-1.2$. The \limr\ stars are found primarily between \feh\ $ =-2.8$ and \feh =$-1.5$. 
Our \rI\ stars cover a wider range of metallicities than those of the RPA-1 and RPA-2 samples, with one \rI\ star identified at \feh\ $ = -3.4$, and another at \feh $ = -1.2$. A similar cutoff in the \rI\ stars distribution is found around \feh$=-3.0$.
The mean metallicity of our newly identified \rII\ stars is \feh\ $\sim-2.3$, which agrees with the RPA-2 results. The \feh\ distribution from the RPA-1 sample is slightly shifted toward more metal-poor values, which can be explained by their implementation of an LTE analysis as compared to the 1D, NLTE and $<$3D$>$, NLTE analyses implemented in this work and RPA-2 papers respectively.
Our sample of stars also shift toward higher V magnitudes than those analyzed by the RPA-1 and RPA-2 efforts (Figure\,\ref{fig:hist}, lower panels). This shift correlates with the sizes of the telescopes used in this work and those by RPA-1 and RPA-2 efforts. Our \rII\ stars are particularly extended up to $V=14.2$ dex. 

We also show the [Eu/Fe] abundance distribution, as a function of metallicity, in Figure~\ref{fig:abund_eufeh}. For comparison, we also show the stars from the RPA-1 and RPA-2 samples, as well as those from literature studies by \citet{westin2000}, \citet{johnson2002}, \citet{cayrel2004}, \citet{honda2004}, \citet{barklem2005}, \citet{ivans2006}, \citet{mashonkina2010}, \citet{SCM2014}, and \citet{roederer2014b}. The histograms on the top and right axes represent the number of stars covering each abundance range. Our results broadly overlap with the literature values, where the [Eu/Fe] ratios also cover a wide range of metallicities. 
This reiterates the interpretation by \citet{sneden2000}, who suggested multiple formation sites for \rproc-enhanced stars and a need for investigating \rproc\ production and its site(s). We also plot the absolute Eu abundances derived for our target stars, as a function of \feh\, in Figure\,\ref{fig:eu_abs}. A clear increase of Eu with Fe is seen for the \rII, \rI, \limr, and non-RPE stars. A similar trend was found by the RPA-1 survey \citep{rpa1}.

We additionally show the [Ba/Eu] ratios, as a function of \feh, in Figure\,\ref{fig:abund_baeu}. This can be used as a diagnostic for the level of $r$- versus $s$-process origin of the neutron-capture
elements measured in stars \citep{frebel-rev2018}. We also plot the  RPA-1 and RPA-2 stars on the same plot for comparison. Just as in previous work, the majority of newly identified \rII\ stars strictly have [Ba/Eu] $ < -0.5$, indicating a pure \rproc\ origin, while the \rI\ and \limr\ stars show larger variations in [Ba/Eu]. 

\subsection{Other Elemental Abundances}\label{sec:elem_abund}

The abundance ratios of the light elements, $\alpha$-elements, and Fe-peak elements relative to Fe are shown in Figure\,\ref{fig:abund_rel}, compared with literature values from \citet{cayrel2004}, \citet{barklem2005}, and \citet{cohen2013}. Classes of \rII, \rI, \limr, and non-RPE star are distinguished to investigate possible systematic differences. 

[C/Fe] and [O/Fe] ratios exhibit increasing trends toward lower metallicities.  [Mg/Fe], [Si/Fe], and [Ca/Fe] ratios show typical [$\alpha$/Fe] enhancements, of $\sim +0.4$, on average, while the [Al/Fe] ratios exhibit sub-solar values, except for one \limr\ and four \rI\ stars. We note that [Ti/Fe] ratios  derived from the \tiii\ lines are shown, as they are less prone to NLTE effects than \tii\ lines \citep{sitnova2016}.

Fe-peak element abundance ratios mostly agree with those from the literature.  [Co/Fe] and [Zn/Fe] abundances exhibit increasing trends toward lower metallicities, in agreement with literature results (e.g., \citealt{Ezzeddine2019}), whereas [Mn/Fe] abundances show decreasing trends toward lower metallicities.

The abundances of the majority of our stars agree with typical Galactic field metal-poor star abundances, with the exception of the \rII\ star 2MASS~J00101758$-$1735387 (shown by black star symbol in Figure\,\ref{fig:abund_rel}), 
which exhibits sub-solar and lower abundance ratios than the rest of the sample stars for C, Na, Mg,  Al, K, Ca, Sc, Ti, V, Cr, Mn, Co, Ni, and Zn. This star exhibits similar abundance patterns to other halo \rI\ and \rII\ stars identified in \citet{sakari2019} and \citet{xing2019}. These stars exhibit similar abundance patters to those identified in Milky Way (MW) dwarf galaxies \citep{mcwilliam2018}. The former studies suggested that such halo stars have an accretion origin from a dwarf galaxy previously enriched by a NS merger event, which enriched the gas from which 2MASS~J00101758$-$1735387 was formed. Additional investigation of its full \rproc\ abundance pattern is underway, and will be presented in a future paper. 

Another \rII\ star, \mbox{2MASS~J05241392$-$0336543} 
exhibits a substantially enhanced [Na/Fe] ratio ($+$1.46).
As shown in Figure~\ref{fig:abund_rel} by the yellow circle marker, the [Mg/Fe] ratio in this star ([Mg/Fe]~$= +$0.09)
is slightly lower than most other stars in the sample, but all other light-element abundance ratios in \mbox{2MASS~J05241392$-$0336543} are typical
for its metallicity (\feh $= -2.50$). 

The atypical [Na/Fe] and [Mg/Fe] abundance ratios appear unrelated to the light-element abundance variations typically observed in globular cluster stars, and discussion of its abundance anomalies and their origin is deferred to a future paper.

\section{Summary and Conclusions}\label{sec:conc}

In this paper we have presented high-resolution spectroscopic observations for 148 newly studied metal-poor stars. We identify 32 new stars with $-2.0\leq$\,[Fe/H]\,$<-1.0$, 103  stars with $-3.0\leq$\,[Fe/H]\,$<-2.0$, 5 stars with $-4.0\leq$\,[Fe/H]\,$<-3.0$, as well as one new ultra metal-poor star at [Fe/H]\,$<-4.0$. One of the stars is a rediscovery of the  previously observed ultra metal-poor star,  CD$-38\,^{\circ}245$. This is part of the ongoing RPA effort to identify new \rproc-enhanced stars in the Galaxy. Seven stars were found to be double-lined spectroscopic binaries, and were thus excluded from further study, leaving the sample to contain 141 newly studied stars.

We determine 1D, NLTE atmospheric stellar parameters for all stars, and derive abundances for the light elements, including C and O, the $\alpha$-elements, the Fe-peak elements, and the neutron-capture elements, whenever suitable spectral lines are available. Among the 141 stars, we find 10 new \rII, 62 new \rI, and 17 new \limr\ stars, based on their Eu, Ba, and Sr abundances. Additionally, we identified one new \sproc-enhanced star and five new \rs\ stars.

Of the ten \rII\ stars, one is a CEMP star,  another is found to exhibit sub-solar elemental abundance ratios relative to Fe for many of the light, $\alpha$, and Fe-peak elements, consistent with accretion from a dwarf galaxy. All of the \rII\ stars exhibit consistently low ratios of [Ba/Eu] ($<-0.5$), which is a signature of pure \rproc\ enhancement.
The \rII\ and \rI\ stars span a range of metallicities, from $-3.0$ to $-1.5$ in [Fe/H], with one \rI\ star detected at \feh\,$=-3.4$ and another at \feh\,$=-1.2$.

The neutron-capture and other elemental abundance ratios were compared to literature values, including the results from the RPA-1 and RPA-2 pilot samples of \citet{rpa1} and \citet{rpa2}, respectively. The metallicity distribution of our stars derived using 1D, NLTE agrees with those derived for the RPA-2 sample \citep{rpa2} using $<$3D$>$, NLTE. The RPA-1 \feh\ distribution, determined using 1D, LTE, is slightly shifted to the metal-poor end as compared to our sample, due to NLTE effects. Our sample is also shifted toward less bright stars as compared to the RPA-1 and RPA-2 sample. Our \rII\ stars cover a range of magnitudes, centered roughly around $V=13$.
The results from the three RPA releases from the Southern Hemisphere (this work, and \citealp{rpa1}) and the Northern Hemisphere \citep{rpa2} have significantly increased the numbers of known \rII, \rI, and \limr\ stars, roughly doubling the previously known samples to
a total of 28 \rII\ stars, 163 \rI\ stars, and 56 \limr\ stars, as well as 6 stars enhanced at the \rs\ level.
The RPA has already obtained snapshot spectroscopy for $\sim 1500$ bright metal-poor stars, and new snapshot spectra for $\sim 1000$ additional candidate stars are presently being obtained with a variety of telescopes in the Northern and Southern Hemispheres.  These observations are planned to be publicly released, as the data are reduced and analyzed, over the course of the next several years. 

The full sample obtained by the RPA will be used to establish what fraction of accreted dwarf galaxies may have contributed metal-poor stars to the halo of the Milky Way; some partial information is already feasible to obtain \citep{roederer2018}.  In addition, detailed information on the nature of the environment in which the $r$-process events associated with individual stars occurred is being explored by consideration of the association of \rproc-enhanced stars with the so-called ``Dynamically Tagged Groups" (DTGs) of very metal-poor stars (Yuan et al., submitted; Gudin et al., in preparation), following the pioneering work of \citet{roederer2018}. When ultimately combined with new accelerator results on the fundamental nuclear properties for isotopes specific to the \rproc\ (using, e.g., the Facility for Rare Isotope Beams; FRIB, expected 2022 \citealt{FRIB2019}),
stellar abundances will provide key missing information required to answer many of the outstanding questions regarding the nature and astrophysical sites of the \rproc.


\acknowledgements 
We thank Gary Da Costa for providing the SkyMapper targets.
R.E., A.F., V.M.P, T.C.B., and I.U.R. acknowledge support from JINA-CEE (Joint Institute for Nuclear Astrophysics - Center for the Evolution of the Elements), funded by the NSF under Grant No. PHY-1430152. I.U.R. acknowledges partial support from NSF grants AST-1613536 and AST-1815403. A.F. is partially supported by NSF-CAREER grant AST-1255160 and NSF grant 1716251. JM thanks FAPESP (2014/18100-4).
This work made use of NASA's Astrophysics Data System Bibliographic Services, and the SIMBAD database, operated at CDS, Strasbourg, France \citep{simbad}. This work has made use of data from the European Space Agency (ESA) mission
{\it Gaia} (\url{https://www.cosmos.esa.int/gaia}), processed by the {\it Gaia}
Data Processing and Analysis Consortium (DPAC,
\url{https://www.cosmos.esa.int/web/gaia/dpac/consortium}). Funding for the DPAC
has been provided by national institutions, in particular the institutions
participating in the {\it Gaia} Multilateral Agreement.

\facilities{Magellan-Clay (MIKE, \citealt{bernstein2003})}

\software{IRAF~\citep{iraf}, matplotlib~\citep{matplotlib}, MOOG~\citep{sneden1973,sobeck2011}, MULTI~\citep{Carlsson1986,carlsson1992}, MARCS~\citep{gustafsson1975,gustafsson2008}}
 
\bibliography{ref}

\begin{longrotatetable}
\begin{deluxetable*}{l c c c c c c c r r c}
\tablewidth{0pt}
\tabletypesize{\footnotesize}
\tablecaption{\label{tab:ident} Observational Details of the Target Stars}
\tablehead{
\colhead{Star ID} & \colhead{R.A.} & \colhead{Decl.} & \colhead{$V$ mag\tablenotemark{a}} & \colhead{Exp. Time} & \colhead{Slit} & \colhead{MJD} & \colhead{S/N}  & \colhead{RV$_\mathrm{helio}$} & \colhead{RV$_\mathrm{helio}$(Gaia)} & \colhead{Source}\\
&  \multicolumn{2}{c}{(J2000)}   &          & \colhead{(sec)} & \colhead{(arcsec)} & & \colhead{(@4000{\AA})}  & \colhead{(\kms)} & \colhead{(\kms)} & }
\startdata
2MASS~J00011195$+$0321051  &   00:01:12.000   &  $+$03:21:03.600  &  12.802  &   1500   &   1.0    & 58034.5 &  60  &$-$211.75&  $-$208.14&    RAVE              \\
2MASS~J00101758$-$1735387  &   00:10:17.520   &  $-$17:35:38.400  &  11.539  &   1200   &   0.7    & 56940.5 &  100 & 191.80 &    191.76&    M\&P               \\
2MASS~J00162809$-$0505519  &   00:16:28.100   &  $-$05:05:52.100  &  12.921  &   1200   &   0.7    & 58385.5 &  65  & $-$168.57& $-$173.48&    SkyMapper         \\
2MASS~J00195350$-$0145445  &   00:19:53.520   &  $-$01:45:43.200  &  12.838  &   1800   &   1.0    & 58034.5 &  50  & $-$301.34& $-$321.93&    RAVE              \\
2MASS~J00282278$-$3934598  &   00:28:22.800   &  $-$39:34:59.900  &  12.547  &   1200   &   1.0    & 58380.5 &  60  & 29.20    &   23.56&    SkyMapper         \\
2MASS~J00372794$-$2045307  &   00:37:27.930   &  $-$20:45:30.640  &  12.642  &   1200   &   0.7    & 58385.5 &  65  & 68.11    &   64.14&    RAVE              \\
2MASS~J00452379$-$2112161  &   00:45:23.800   &  $-$21:12:16.100  &  12.538  &   1200   &   0.7    & 58385.5 &  75  & 26.15    &   21.19&    SkyMapper         \\
2MASS~J00463619$-$3739335\tablenotemark{b}&00:46:36.190&$-$37:39:33.550&11.929&  1200   &   1.0    & 58362.5 &  60  & 42.70    &   48.28&    RAVE              \\
\enddata
\tablenotetext{a}{Magnitudes from RAVE DR5 \citep{kunder2017}.}
\tablenotetext{b}{Re-discovery of the ultra-metal-poor (UMP) star CD-38$^{o}$245.}
\tablenotetext{c}{Double-lined spectroscopic binaries (SB2).}
\end{deluxetable*}
\end{longrotatetable}

\startlongtable
\begin{deluxetable*}{l r l r r r r}
\tablewidth{0pt}
\tabletypesize{\footnotesize}
\tablecaption{\label{tab:linelist} Atomic line properties, equivalent widths and abundances of the target stars. Only one star is shown here. Full table for the rest of sample will be available electronically.}
\tablehead{
\colhead{Star ID} & \colhead{$\lambda$} & \colhead{Species} & \colhead{$\chi$} & \colhead{$\log\,gf$} & \colhead{EW} & \colhead{$A(X)$}\\
& \colhead{({\AA})}     &  & \colhead{(eV)} &     &\colhead{(m{\AA})}    & \colhead{(dex)}} 
\startdata
2MASS~J00011195$+$0321051 &    5688.205 &      \nai\ &      2.10 &  $-$0.452 &     15.09 &      4.22 \\
2MASS~J00011195$+$0321051 &    5895.924 &      \nai\ &      0.00 &  $-$0.194 &    175.60 &      4.28 \\
2MASS~J00011195$+$0321051 &    4057.505 &      \mgi\ &      4.35 &  $-$0.890 &     92.51 &      6.24 \\
2MASS~J00011195$+$0321051 &    4167.271 &      \mgi\ &      4.35 &  $-$0.710 &     88.96 &      6.00 \\
2MASS~J00011195$+$0321051 &    4571.096 &      \mgi\ &      0.00 &  $-$5.688 &     98.88 &      6.06 \\
2MASS~J00011195$+$0321051 &    4702.990 &      \mgi\ &      4.33 &  $-$0.380 &    107.99 &      5.81 \\
2MASS~J00011195$+$0321051 &    5528.405 &      \mgi\ &      4.34 &  $-$0.498 &    113.85 &      6.00 \\
2MASS~J00011195$+$0321051 &    5711.090 &      \mgi\ &      4.34 &  $-$1.724 &     30.35 &      5.88 \\
\enddata
\end{deluxetable*}

\begin{longrotatetable}
\begin{deluxetable*}{l c c c c c c c c c c c c c c c c c c c}
\tablewidth{0pt}
\tabletypesize{\scriptsize}
\tablecaption{\label{tab:stell_param} Stellar Atmospheric Parameters of the Target Stars}
\tablehead{
\\
& \multicolumn{4}{c}{LTE}   & & \multicolumn{5}{c}{LTE$^{\mathrm{corr}}$} & & \multicolumn{5}{c}{NLTE} \\
\cline{2-5} \cline{7-11} \cline{13-17}
\colhead{Star ID} & \colhead{\teff} & \colhead{\logg} & \colhead{\vt} & \colhead{\feh} & & \colhead{\teff} & \colhead{\logg} & \colhead{\vt}  & \colhead{\feh} & \colhead{$\sigma_{\mathrm{[Fe/H]}}$} & & \colhead{\teff} & \colhead{\logg} & \colhead{\vt}  & \colhead{\feh} & \colhead{$\sigma_{\mathrm{[Fe/H]}}$} & \colhead{N$_\mathrm{\fei}$} & \colhead{N$_\mathrm{\feii}$} & \colhead{\logg$^{\mathrm{Gaia}}$} \\
 & \colhead{(K)} & \colhead{(cgs)} & \colhead{(\kms)} & & & \colhead{(K)} & \colhead{(cgs)} & \colhead{(\kms)}  &  & \colhead{(dex)} & & \colhead{(K)} & \colhead{(cgs)} & \colhead{(\kms)}  &  & \colhead{(dex)} &  & & \colhead{(cgs)}
}
\startdata
 2MASS~J00011195$+$0321051  &  4710  &  1.25  &  1.80  & $-$2.04 & &  4909  &  1.75  &  1.85  & $-$1.99  &	0.12  &  &  4820  &    2.20	 & 1.80	 &  $-$1.76 &  0.12  & 237  &  26  & 2.10  \\
 2MASS~J00101758$-$1735387  &  5000  &  1.40  &  2.10  & $-$2.64 & &  5229  &  2.75  &  1.50  & $-$2.41  &	0.12  &  &  5200  &    3.00	 & 1.70	 &  $-$2.19 &  0.12  & 262  &  37  & 2.91  \\
 2MASS~J00162809$-$0505519  &  4865  &  1.50  &  1.75  & $-$3.23 & &  5049  &  1.90  &  1.70  & $-$3.07  &	0.10  &  &  4920  &    2.45	 & 1.75	 &  $-$2.82 &  0.10  &  179  &  18  &2.50  \\
 2MASS~J00195350$-$0145445  &  4590  &  0.60  &  2.45  & $-$2.67 & &  4801  &  1.18  &  2.29  & $-$2.45  &	0.13  &  &  4720  &    1.95	 & 2.10	 &  $-$2.22 &  0.13  & 216  &  24  & 1.91  \\
 2MASS~J00282278$-$3934598  &  4440  &  0.40  &  2.75  & $-$2.88 & &  4666  &  1.10  &  2.55  & $-$2.61  &	0.13  &  &  4550  &    1.45	 & 2.40	 &  $-$2.36 &  0.13  & 194  &  20  & 1.38  \\
 2MASS~J00372794$-$2045307  &  4615  &  0.95  &  2.25  & $-$2.80 & &  4824  &  1.50  &  2.05  & $-$2.64  &	0.08  &  &  4750  &    2.00	 & 2.10	 &  $-$2.40 &  0.08  & 220  &  24  & 1.89  \\
 2MASS~J00452379$-$2112161  &  4570  &  0.60  &  2.15  & $-$2.97 & &  4783  &  1.25  &  2.05  & $-$2.63  &	0.14  &  &  4650  &    1.85	 & 2.10	 &  $-$2.40 &  0.14  & 247  &  26  & 1.79  \\
 2MASS~J00463619$-$3739335  &  4525  &  0.60  &  2.80  & $-$4.26 & &  4743  &  1.20  &  2.00  & $-$4.12  &	0.23  &  &  4650  &    1.50	 & 2.00	 &  $-$4.02 &  0.23  & 67  &  6    & 1.44  \\
\enddata
\tablenotetext{a}{Uncertain \logg$^{\mathrm{Gaia}}$ due to $Gaia$ DR2 parallax-based distance larger than 6\,kpc \citep{bailer-jones2018}.}
\end{deluxetable*}
\end{longrotatetable}

\startlongtable
\begin{deluxetable*}{l r r r r r r r c}
\tablewidth{0pt}
\tabletypesize{\footnotesize}
\tablecaption{\label{tab:rproc_clas} Stellar Sub-classifications and Associated Abundances}
\tablehead{
\colhead{Star ID} & \colhead{{  }{  }{  }$\mathrm{[C/Fe]}$} & \colhead{$\mathrm{[C/Fe]}_\mathrm{corr.}$} & \colhead{$\mathrm{[Sr/Fe]}$} & \colhead{$\mathrm{[Ba/Fe]}$} & \colhead{$\mathrm{[Eu/Fe]}$} & \colhead{$\mathrm{[Ba/Eu]}$} & \colhead{$\mathrm{[Sr/Ba]}$} & Sub-Class}
\startdata
2MASS~J00011195$+$0321051 &     0.00 &     0.22  &      0.20  &     0.20 &     0.22 &     $-$0.02  &     0.00  &   non-RPE           \\            
2MASS~J00101758$-$1735387 &     0.00 &     0.01  &      0.09  &     0.43 &     1.35 &     $-$0.92  &  $-$0.34  &   $r$-II            \\            
2MASS~J00162809$-$0505519 &     0.60 &     0.66  &      0.01  &  $-$1.21 &  $<$0.10 &    $>-$1.31  &     1.22  &    limited-$r$?      \\            
2MASS~J00195350$-$0145445 &  $-$0.45 &     0.23  &   $-$0.32  &  $-$0.24 &     0.28 &     $-$0.52  &  $-$0.08  &   non-RPE           \\            
2MASS~J00282278$-$3934598 &  $-$0.50 &     0.23  &   $-$0.06  &  $-$0.40 &  $-$0.20 &     $-$0.20  &     0.34  &    non-RPE          \\            
2MASS~J00372794$-$2045307 &     0.20 &     0.63  &   $-$0.03  &  $-$0.39 &  $-$0.28 &     $-$0.11  &     0.36  &    non-RPE          \\            
2MASS~J00452379$-$2112161\tablenotemark{a} &  $-$0.15 &     0.50  &   $-$0.10  &  $-$0.25 &     0.40 &     $-$0.65  &     0.15  &    $r$-I \\      
2MASS~J00463619$-$3739335 & $<-$3.31 & $<-$2.72  &   $-$0.64  &  $-$0.59 &  $<$0.80 &    $>-$1.39  &  $-$0.05  &    non-RPE          \\
\enddata
\tablenotetext{a}{Rediscovery of the ultra metal-poor (UMP) star CD$-38+245$ }
\end{deluxetable*}

\begin{longrotatetable}
\begin{deluxetable*}{l r r r r r r r r c}
\tablewidth{0pt}
\tabletypesize{\footnotesize}
\tablecaption{\label{tab:abund_light} Light-Element and $\alpha$-Element abundances}
\tablehead{
\colhead{Star ID} & \colhead{$\mathrm{[O/H]}$} & \colhead{$\mathrm{[Na/H]}$} & \colhead{$\mathrm{[Mg/H]}$} &
\colhead{$\mathrm{[Al/H]}$} & \colhead{$\mathrm{[Si/H]}$} & \colhead{$\mathrm{[K/H]}$} & \colhead{$\mathrm{[Ca/H]}$} & \colhead{$\mathrm{[\tii/H]}$} & \colhead{$\mathrm{[\tiii/H]}$}
}
\startdata
2MASS~J00011195$+$0321051     &    $-0.79\pm 0.10$   &    $-1.99\pm0.03$  &     $-1.60\pm0.14$ &   $-2.33\pm0.15$  &   $-1.83\pm0.10$     &     $  -1.33\pm	0.10 $   &  $-1.65\pm0.14$   & $-1.88\pm0.14$ &  $-1.76\pm0.17$   \\
2MASS~J00101758$-$1735387     &    \nodata          &     $-3.01\pm0.08$  &     $-2.75\pm0.08$ &   $-3.73\pm0.15$  &   $-2.14\pm0.10$     &     $  -2.73\pm	0.10 $   &  $-2.19\pm0.08$   & $-2.67\pm0.17$ &  $-2.57\pm0.22$   \\
2MASS~J00162809$-$0505519     &    \nodata           &    $-2.72\pm0.10$  &     $-2.52\pm0.07$ &   $-3.57\pm0.15$  &   $-2.61\pm0.10$     &     $  -2.67\pm	0.10 $   &  $-2.66\pm0.11$   & $-2.81\pm0.14$ &  $-2.81\pm0.14$   \\
2MASS~J00195350$-$0145445     &    $-1.59\pm 0.04$   &    $-2.30\pm0.20$  &     $-2.02\pm0.08$ &   $-2.65\pm0.15$  &   $-1.85\pm0.09$     &     $  -1.66\pm	0.13 $   &  $-2.09\pm0.14$   & $-2.30\pm0.08$ &  $-2.13\pm0.18$   \\
2MASS~J00282278$-$3934598     &    $-1.56\pm0.10  $  &    $-2.49\pm0.02$  &     $-2.18\pm0.05$ &   $-3.25\pm0.15$  &   $-1.93\pm0.10$     &     $  -2.15\pm	0.10 $   &  $-2.28\pm0.08$   & $-2.37\pm0.13$ &  $-2.17\pm0.15$   \\
2MASS~J00372794$-$2045307     &    \nodata           &    $-2.51\pm0.01$  &     $-2.13\pm0.07$ &   $-3.18\pm0.15$  &   $-2.02\pm0.05$     &     $  -2.19\pm	0.00 $   &  $-2.12\pm0.08$   & $-2.50\pm0.11$ &  $-2.42\pm0.14$   \\
2MASS~J00452379$-$2112161     &    $-2.12\pm0.10  $  &    $-1.92\pm0.18$  &     $-2.28\pm0.06$ &   $-2.32\pm0.15$  &   $-1.94\pm0.17$     &     $  -1.92\pm	0.10 $   &  $-2.22\pm0.10$   & $-2.38\pm0.09$ &  $-2.32\pm0.12$   \\
2MASS~J00463619$-$3739335     &   \nodata            &    $-4.10\pm0.10$  &     $-3.52\pm0.13$ &   $-4.67\pm0.15$  &   $-3.59\pm0.10$     &     \nodata               &  $-3.66\pm0.10$   & $-3.04\pm0.20$ &  $-3.66\pm0.25$   \\
\enddata
\end{deluxetable*}
\end{longrotatetable}

\begin{longrotatetable}
\begin{deluxetable*}{l r r r r r r r}
\tablewidth{0pt}
\tabletypesize{\footnotesize}
\tablecaption{\label{tab:abund_fe_peak}Fe-Peak Element Abundances}
\tablehead{
\\
\colhead{Star ID} & \colhead{$\mathrm{[Sc/H]}$} & \colhead{$\mathrm{[V/H]}$} & \colhead{$\mathrm{[Cr/H]}$} & \colhead{$\mathrm{[Mn/H]}$}& \colhead{$\mathrm{[Co/H]}$} & \colhead{$\mathrm{[Ni/H]}$} & \colhead{$\mathrm{[Zn/H]}$}
}
\startdata
2MASS~J00011195$+$0321051     &  $-1.97\pm0.12$	 &  $-1.72\pm0.17$ &  $-1.88\pm0.04$  &  $-2.35\pm0.12$   &  $-1.92\pm0.12$  &  $ -2.17\pm0.16$   &  $-1.87\pm0.05$  \\
2MASS~J00101758$-$1735387     &  $-3.20\pm0.18$	 &  $-2.56\pm0.17$ &  $-2.25\pm0.07$  &  $-3.08\pm0.07$   &  $-2.66\pm0.20$  &  $ -3.02\pm0.00$   &  $-3.09\pm0.00$  \\
2MASS~J00162809$-$0505519     &  $-3.04\pm0.12$	 &  $-2.48\pm0.19$ &  $-3.17\pm0.00$  &  $-3.43\pm0.11$   &  $-2.74\pm0.11$  &  $ -2.82\pm0.14$   &  $-2.72\pm0.00$  \\
2MASS~J00195350$-$0145445     &  $-2.35\pm0.10$	 &  $-2.24\pm0.09$ &  $-2.40\pm0.07$  &  $-2.84\pm0.08$   &  $-2.42\pm0.18$  &  $ -2.45\pm0.16$   &  $-2.28\pm0.04$  \\
2MASS~J00282278$-$3934598     &  $-2.47\pm0.12$	 &  $-2.07\pm0.14$ &  $-2.37\pm0.01$  &  $-2.99\pm0.11$   &  $-2.50\pm0.20$  &  $ -2.46\pm0.12$   &  $-2.27\pm0.06$  \\
2MASS~J00372794$-$2045307     &  $-2.66\pm0.08$	 &  $-2.37\pm0.08$ &  $-2.53\pm0.02$  &  $-2.66\pm0.10$   &  $-2.44\pm0.13$  &  $ -2.53\pm0.13$   &  $-2.22\pm0.07$  \\
2MASS~J00452379$-$2112161     &  $-2.51\pm0.09$	 &  $-2.38\pm0.10$ &  $-2.49\pm0.02$  &  $-2.98\pm0.12$   &  $-2.42\pm0.15$  &  $ -2.55\pm0.13$   &  $-2.43\pm0.05$  \\
2MASS~J00463619$-$3739335     &  $-3.92\pm0.16$	 &  \nodata        &  $-3.08\pm0.00$  &  $-4.77\pm0.29$   &  $-3.44\pm0.17$  &  $ -3.65\pm0.23$   &  $-3.38\pm0.02$  \\
\enddata
\end{deluxetable*}
\end{longrotatetable}

\end{document}